\documentclass[12pt,amsmath,amssymb,unsortedaddress,superscriptaddress,prb]{revtex4-1}

\usepackage{subeqn}
\usepackage{graphicx}
\usepackage{dcolumn}
\usepackage{bm}
\usepackage{pifont}

\begin{document}


\title{Dopant-enhanced solid phase epitaxy in buried amorphous silicon layers}
\author{B. C. Johnson\footnote{Email: bcj@physics.unimelb.edu.au}}
\author{J. C. McCallum}%
\affiliation{%
School of Physics, University of Melbourne, Victoria 3010, Australia
}%


\begin{abstract}
The kinetics of intrinsic and dopant-enhanced solid phase epitaxy (SPE)
are studied in buried amorphous Si ($a$-Si) layers in which SPE
is not retarded by H. As, P, B and Al profiles were formed by multiple energy ion implantation over a concentration range of $\rm 1 - 30 \times 10^{19} \; /cm^{3}$. Anneals were performed in air over the temperature range 460-660 $\rm ^{o}$C and the rate of interface motion was monitored using time resolved reflectivity. The dopant-enhanced SPE rates were modeled with the generalized Fermi level shifting model using degenerate semiconductor statistics. The effect of band bending between the crystalline and amorphous sides of the interface is also considered.

\end{abstract}

\maketitle

\section{Introduction}
Solid phase epitaxy (SPE) is an important processing step in device
fabrication yet the mechanism by which atoms make the amorphous to
crystalline phase transition is still poorly understood. It is known
that the velocity of the crystalline-amorphous ($c$-$a$) interface
(SPE rate) has a strong dependence on a number of parameters including
pressure,\cite{APL:lu,JAP:lu} substrate orientation,\cite{JAP:csepregi2}
and dopant concentration\cite{JAP:csepregi}. The SPE rate in intrinsic
silicon is also well described by an Arrhenius type equation with an
activation energy of 2.7~eV.\cite{APL:roth}

Pioneering work on dopant-enhanced SPE was performed by Csepregi and
co-workers who used ion channeling measurements to determine surface
$a$-Si layer thicknesses as a function of annealing time and
temperature.\cite{JAP:csepregi} Csepregi found that the presence of
impurity atoms enhanced the SPE rate up to 6 times that of the
intrinsic rate for P and As and up to 20 times for B with impurity
concentrations of 0.4 at.\%. Further, it has been found that
overlapping $n$- and $p$-type dopants of a similar concentration give
an SPE rate close to that of an intrinsic sample, a compensation
doping effect.\cite{APL:suni} Likewise, the overlapping of two dopants
of the same type gives an additive effect on the SPE regrowth
rate. These electronic effects suggest that the SPE rate is sensitive
to shifts in the Fermi level and that both neutral and charged defects
may be responsible for the SPE process. Indeed, a number of models link the structural and electrical properties at the $c$-$a$ interface in an attempt to describe the SPE phenomenon.
\cite{PRL:williams,APL:suni,JAP:lietoila,JAP:mosley,MRSSP:walser,APL:park,JAP:lu} However, there is a significant problem with the experimental data for SPE growth of surface amorphous layers to which the SPE models have been fitted.
In surface amorphous layers, the SPE rate is invariably affected by the
presence of H which infiltrates the layer from the native oxide and retards
the SPE growth rate.\cite{APL:roth}  Roth {\em et al.} have shown that the
growth rate is affected for interface depths of up to $\sim \rm 1.5\: \mu m$. Recent measurements also suggest that the infiltration of H can affect 
the dopant-enhanced SPE regrowth rate.\cite{tobespehdas} These findings call
into question the reliability of the parameters found in previous work where
SPE models have been fitted to the H-affected data.

This paper presents data for dopant-enhanced SPE kinetics in 
buried $a$-Si layers. Buried $a$-Si layers provide an environment where H
concentrations at the $c$-$a$ interface are considerably lower than the
levels which have a measurable affect on SPE rates.\cite{APL:mccallum} 
H-free dopant-enhanced SPE
data are presented for As, P, B and Al measured over a concentration
range of $\rm 1 - 30 \times 10^{19} \; cm^{-3}$. As-enhanced SPE data
are care of McCallum.~\cite{NIMB:mccallum} The experimental data are fitted
to the generalized Fermi level shifting (GFLS) model which is one of the
more highly developed models describing the dopant-enhancement effect in SPE
which at this stage has not met with any significiant
challenge to its validity.~\cite{JAP:lu} By fitting the model to SPE data for
both $n-$ and $p-$type impurities over a broad concentration range we are
able to show that the model provides plausible values for fitting
parameters which may eventually lead to the identification of
the defect(s) responsible for the SPE process. In fitting the data, we
have refined the GFLS model by incorporating degenerate semiconductor
statistics and by endeavoring to use the most valid values and dependencies for parameters which affect semiconductor statistics. These parameters include the electron and hole effective masses and the
temperature and concentration dependencies of the band gap. An extension of the
model to include the effect of band bending at the $c$-$a$ interface is
discussed. Further, the scope for developing links between our data and molecular dynamics simulations of the interface motion during SPE as a possible means of identifying the mechanism giving rise to dopant-enhancement are also discussed.

In section \ref{gfls} of this paper we provide a brief overview of the theoretical background to the GFLS model. In section \ref{experiment} we present the experimental parameters associated with our measurements of intrinsic and dopant-enhanced SPE rates. Section \ref{results} presents the SPE rate data and fits to this data using the GFLS model and section \ref{conclusions} presents our conclusions. Details of the parameters and functional dependences used to calculate the semiconductor statistics and hence the Fermi levels in the temperature range of our SPE measurements are given in the appendix.

\section{Theoretical background}
\label{gfls}
Atomistic SPE models have offered some insight into the rate-limiting step of the SPE process and can be used to predict the orientation dependence of the crystallisation rate through the proportionality of
the growth rate with the concentration of [110] ledges. The concentration of these ledges increases with deviations away from the $<\!\!111\!\!>$ crystallographic direction.\cite{JAP:csepregi2} The shortcoming of atomistic approaches lies in their inability to predict the growth rate dependence on dopant concentration which is more likely to be explained by models based on electronic processes.

While conducting experiments on the compensation effect in the
epitaxial regrowth rate of $a$-Si, Suni {\em et al.} suggested that
the bond-breaking process was mediated by vacancies that formed and
migrated at the $c$-$a$ interface.\cite{APL:suni,TSF:suni} This
assumption was based on a finding by Csepregi that the activation
energies of SPE regrowth and the formation of vacancies were
similar.\cite{SSC:csepregi} Using the vacancy model by Van Vechten and
Thurmond,\cite{PRB:vechten} Suni {\em et al.} related the concentration 
of charged vacancies to the position of the Fermi level in the band-gap 
and its dependence on doping concentration. The doping concentration is proportional to the number of charged vacancies while the number of neutral vacancies is not affected. With a greater total concentration of vacancies at the interface the SPE 
growth rate was assumed to be enhanced via some vacancy-related bond-breaking 
process. The assumption that vacancies are the SPE defect has since been 
ruled out due to the observation of a negative activation volume for the 
SPE process in studies where the pressure dependence has been 
measured.\cite{APL:lu} However, the model was significant in that it 
provided a possible mechanism by which the number of growth sites at 
the interface could depend on doping and it related the rate-enhancement 
to the presence of defect energy levels within the band gap. 

The charged kink-site model proposed by Williams and
Elliman,\cite{PRL:williams} is an extension of atomistic models
introduced by Spaepen and Turnbull.\cite{ACTA:spaepen} They 
considered a bond-breaking process involving the propagation of kink-like 
growth sites along [110] ledges. They made reference to defects associated 
with distorted silicon-silicon bonds being responsible
for the bond-breaking process at the kink sites but did not specify the 
exact nature of the defects. Williams and Elliman proposed that the Fermi 
level on the amorphous side of the $c$-$a$ interface would be pinned
near mid-gap and that therefore the number of charged kink-related defects 
at the interface available to promote SPE would be governed by the 
doping dependence of the Fermi-level in the crystalline material. Williams
and Elliman arrived at an expression for the rate-enhancement which is
equivalent to: $v/v_i = N_d.F(T)$, where $v$ is the SPE rate in doped
material, $v_i$ is the rate in intrinsic material, $N_d$ is the dopant
concentration and $F(T)$ is some function of temperature and is independent of $N_d$. In arriving at this expression, they used parameters and
functional dependencies appropriate to an extrinsic non-degenerate 
semiconductor and assumed that the dopants were fully ionized. At the time
the model was proposed, the paucity of reliable SPE velocity data as
a function of dopant concentration meant that the dopant dependence
predicted by the model could not be accurately tested. Since reliable
SPE data which is not affected by H has become available, it is clear that some of the simplifying assumptions which they made are not valid. However, 
their model still has significant merit.

Walser {\em et al.} also introduced a model based on the ideas of Spaepen and Turnbull.\cite{MRSSP:walser,APL:park} The model assumes that the capture of dangling bonds (DB) at the $c$-$a$ interface is the rate-limiting step to the SPE process and that the concentration of these defects is determined by the band structure on the amorphous side of the interface. The Fermi level is assumed to be pinned to mid gap on the amorphous side of the interface and does not affect the SPE rate. Instead, the model suggests that the concentration of DBs is changed by the doping concentration through ionization enhanced atomic mobility as per Bourgoin and Germain.\cite{PL:bourgoin} The equations derived from this model provide reasonable fits to the data. However, the assumptions on which they are based severely limit the model's applicability and interpretation as correctly pointed out by Lu {\em et al.}\cite{JAP:lu} Namely, the law of mass action is violated when assuming that the fractional ionization of dopant atoms in $a$-Si is independent of concentration. It should also be noted that the SPE rate data to which their model was fitted was hydrogen affected.

In their reanalysis of the Williams and Elliman charged kink site model,
Lu {\em et al.} noted that for kink motion to occur, bond
rearrangement must take place and that this would most likely occur via the 
breaking of bonds which span the $c$-$a$ interface followed by local
rearrangement and then recombination of the dangling bonds. Hence, they
considered the kink site model to be a special case of the dangling bond
model of Spaepen and Turnbull\cite{ACTA:spaepen} in which the dangling
bonds are annihilated locally and relatively quickly without taking a
large number of jumps prior to annihilation. Lu {\em et al.} also reworked
the electronic aspects of the charge kink site model, relaxing some of
the assumptions which had been made. They called this reworked model 
the generalized Fermi level shifting (GFLS) model. It is this model which we have further developed by incorporating degenerate semiconductor
statistics and by endeavouring to use the most valid parameters.

In the GFLS model, SPE is thought to occur via bond breaking and 
rearrangement at the $c$-$a$ interface mediated by a neutral defect, $D^{0}$, and its positively or negatively charged counterparts, $D^{\pm}$. 
The defects may be dangling bonds or they may be some other defect. The
model does not attempt to specify this. It is assumed that the defects are 
in thermal and electronic equilibrium and that the concentrations of 
$D^{\pm}$ are determined by the band structure and
density of states (DOS) of the bulk crystal. The SPE regrowth rate is then
expected to be proportional to the concentration of these defects. For
an $n$-type semiconductor and it's intrinsic counterpart the velocities
are given by

\begin{subequations}
\label{velocity}
\begin{equation} v=A([D^{0}]+[D^{-}]|_{\rm doped}) \end{equation} and
\begin{equation} v_{i}=A([D^{0}]+[D^{-}]|_{\rm intrinsic}) \end{equation} 
\end{subequations}

\noindent respectively, where $A$ is a constant and $[D^{0}]$ is the
concentration of neutral defects and is independent of doping. These
equations assume that for $n$-type material, SPE is dominated by $D^{0}$ 
and $D^{-}$ and that each of these defects are equally capable of promoting 
interface motion. The charged fraction of defects is determined by Fermi-Dirac
statistics and, for an $n$-type semiconductor, is expressed as the
ratio of charged to neutral defect concentrations in the crystal,

\begin{equation} 
\label{defectfrac}
\frac{[D^{-}]|_{\rm doped}}{[D^{0}]\;} = g \exp\Bigl(\frac{E_{f}-E^{-}}{kT}\Bigr) 
\end{equation} 

\noindent where $E_{f}$ is the Fermi level and $E^{-}$ represents the
energy level within the band gap of the defect responsible for the
SPE process. The degeneracy factor, $g$, associated with $E^{-}$
is given by $g=Z(D^{-})/Z(D^{0})$ where $Z(D^{-})$ and $Z(D^{0})$ are
the internal degeneracies of the $D^{-}$ and $D^{0}$ defect states,
respectively.\cite{book:bourgoin} If a DB defect is
responsible for the SPE process then it is expected that $g=1/2$ if only
the spin degeneracy needs to be considered. For the positive charge state
of the DB, $g=1$ as the degeneracy of the valence band also contributes
a factor of two.

Once Eq.~\ref{defectfrac} is substituted into the expression for the
velocity we obtain,

\begin{equation} 
\label{fit}
\frac{v}{v_{i}} = \frac{1+\frac{[D^{-}]}{[D^{0}]}|_{\rm doped}}{1+\frac{[D^{-}]}{[D^{0}]}|_{\rm intrinsic}}
 = \frac{1+g\exp\Bigl(\frac{E_{f}-E^{-}}{kT}\Bigr)}{1+g\exp\Bigl(\frac{E_{fi}-E^{-}}{kT}\Bigr)}.
\end{equation} 

\noindent This equation is used to fit the normalized SPE data as a function of temperature with the degeneracy, $g$, and the energy level, $E^{-}$,
of the SPE defect being free parameters. Lu {\em et al.} assumed that
the mobilities of charged and uncharged DBs are
identical.\cite{APL:lu} If the charged and neutral defect
concentrations are weighted separately with a factor $A$ and $A'$ in
Eq.~\ref{velocity} then this will have the affect of weighting the
degeneracy factor in Eq.~\ref{fit} by a value of $A'/A$. The energy
level of the defect predicted by the model will be unaffected by this
assumption.

\begin{figure}[!ht]
\begin{center}
\rotatebox{90}{\includegraphics[height=17cm]{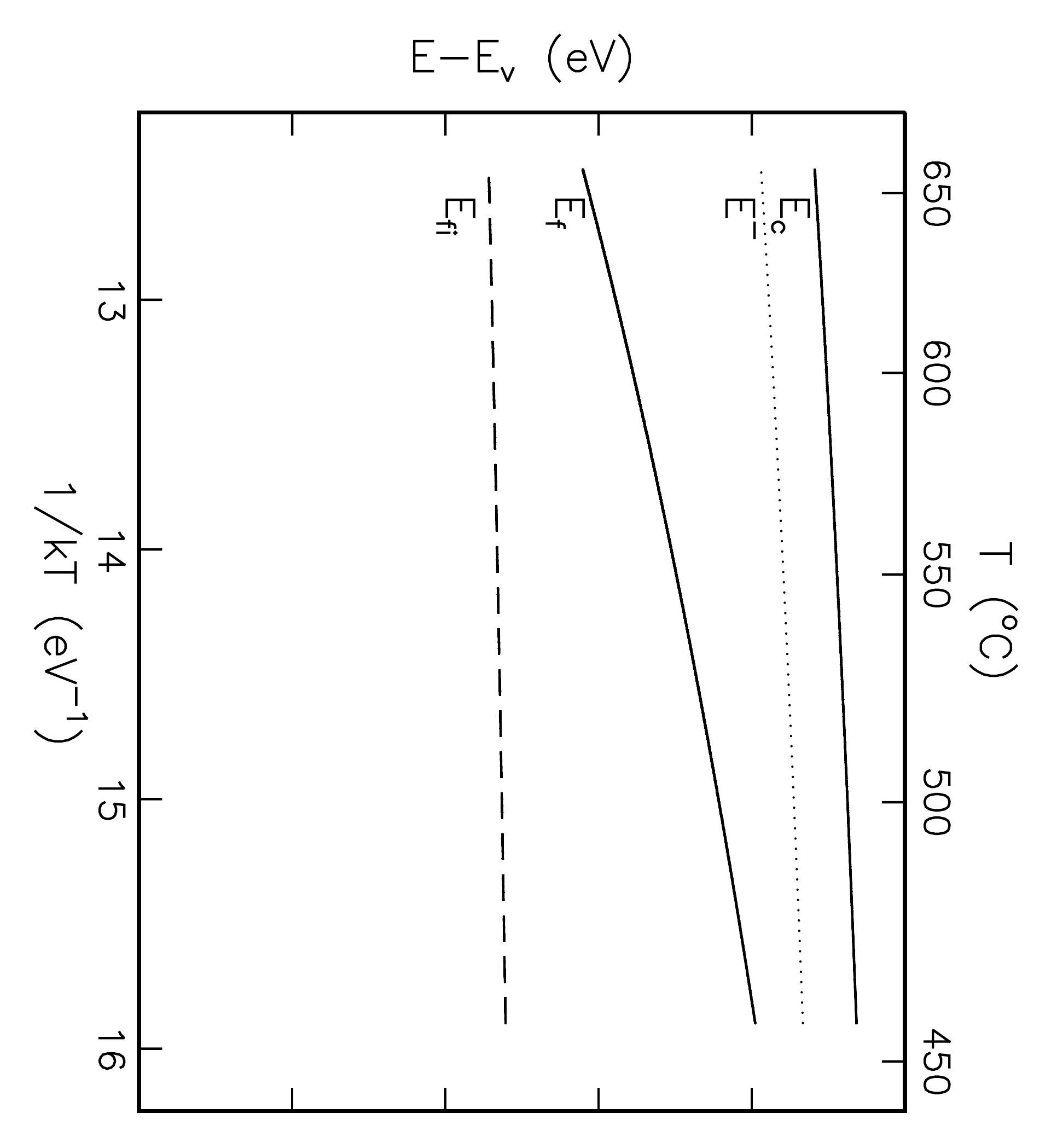}}%
\end{center}
\caption[Fermi levels as a function of temperature]{\label{fermischematic} Schematic of the Fermi levels of an intrinsic, $E_{fi}$ and an $n$-type semiconductor, $E_{f}$ referenced to the valence band edge over the temperature range used in this work. It is assumed that the energy level of the defect responsible for the SPE process, $E^{-}$ tracks the conduction band edge, $E_{c}$.}
\end{figure}

Figure \ref{fermischematic} shows a schematic diagram of the Fermi level in $n$-type Si, $E_{f}$ and intrinsic Si, $E_{fi}$ over the temperature range used in this work. The energy level of the negatively charged state of the SPE defect, $E^{-}$ is assumed to track the conduction band edge, $E_{c}$. The variation of $E_{f}$ with temperature has been exaggerated to show that it approaches the conduction band edge as the temperature decreases. As this occurs, the population of charged SPE defects, $[D^{-}]$ will increase as the $E_{f}$ approaches its energy level. The dopant enhanced SPE rate relative to the intrinsic rate (Eq. \ref{fit}) will then be greatest for anneals performed at lower temperatures. 

A shortcoming of the model is that the temperature dependence of the SPE defect energy level is not known and cannot be included in these calculations as pointed out by Lu {\em et al.}\cite{JAP:lu} However, a detailed account of the model and its application to an extensive data set
should provide much insight into the nature of the SPE process. Two
parameters which must be calculated in order to apply Eq.~\ref{fit} to
normalized SPE data are $E_{fi}$ and $E_{f}$, the Fermi level of an
intrinsic and doped semiconductor, respectively. This calculation and the required semiconductor parameters are outlined in the appendix.

In discussing SPE models it should also be noted that molecular dynamics (MD) simulations are also informative when attempting to identify the SPE mechanism. These simulations describe the structure and rearrangement of atoms at the $c$-$a$ interface during SPE on a microscopic scale. Early models attributed the SPE mechanism to the motion of a dangling bond type defect.\cite{saito81,saito84} This defect aided the rearrangement of atoms at the interface via bond-breaking. More recently, Bernstein {\em et al.} have shown that the SPE may occur through a number of both simple and complex mechanisms.\cite{prb:bernstein98,prb:bernstein00} By using empirical potential simulations they have found that one simple mechanism involves the rotation of two atoms aided by coordination defects which are locally created and annihilated during crystallization. An example of a more complex mechanism involves the migration to the interface of a five-fold coordinated defect which aides the incorporation of two atoms into the crystal matrix. If the MD simulations accurately model the SPE process then doubt is cast on the generally accepted idea that SPE occurs through a single, thermally-activated process. A more complex model would probably then be needed to describe the electronic effects in SPE.

Mattoni {\em et al.} have also described the segregation and precipitation of B during SPE in highly doped Si.\cite{prb:mattoni} This is shown to result in the retardation of the SPE rate and is in agreement with experiment.\cite{review:roth} However, dopant-enhanced SPE is not considered. Indeed, all MD simulations are performed near the melting point of amorphous silicon in order to ensure reasonable simulation times. There are no MD simulations that we know of that have been performed in the temperature range considered in the present work where the effect of the dopants on the SPE rate becomes apparent (as can be seen at the lower temperatures in Fig.~\ref{asdata}-\ref{bdata}). If such MD simulations become possible, dopant-enhanced SPE may be understood to a greater extent on the microscopic level and could then be used to assess the applicability of the GFLS model.

\section{Experiment}
\label{experiment}
The kinetics of dopant-enhanced SPE were measured in buried $a$-Si
layers formed by self-ion-implantation in $n$-type, $\rm 5-10\:
\Omega.cm$, Si(100) Czochralski grown wafers. A 1.7~MV NEC tandem ion
implanter was used for all implants. During implantation, substrates
were tilted $\rm 7^{o}$ off the incident beam axis to avoid channeling
and affixed to the implanter stage with Ag paste to ensure good
thermal contact.

\begin{figure}[!t]
\begin{center}
\rotatebox{90}{\includegraphics[height= 17cm]{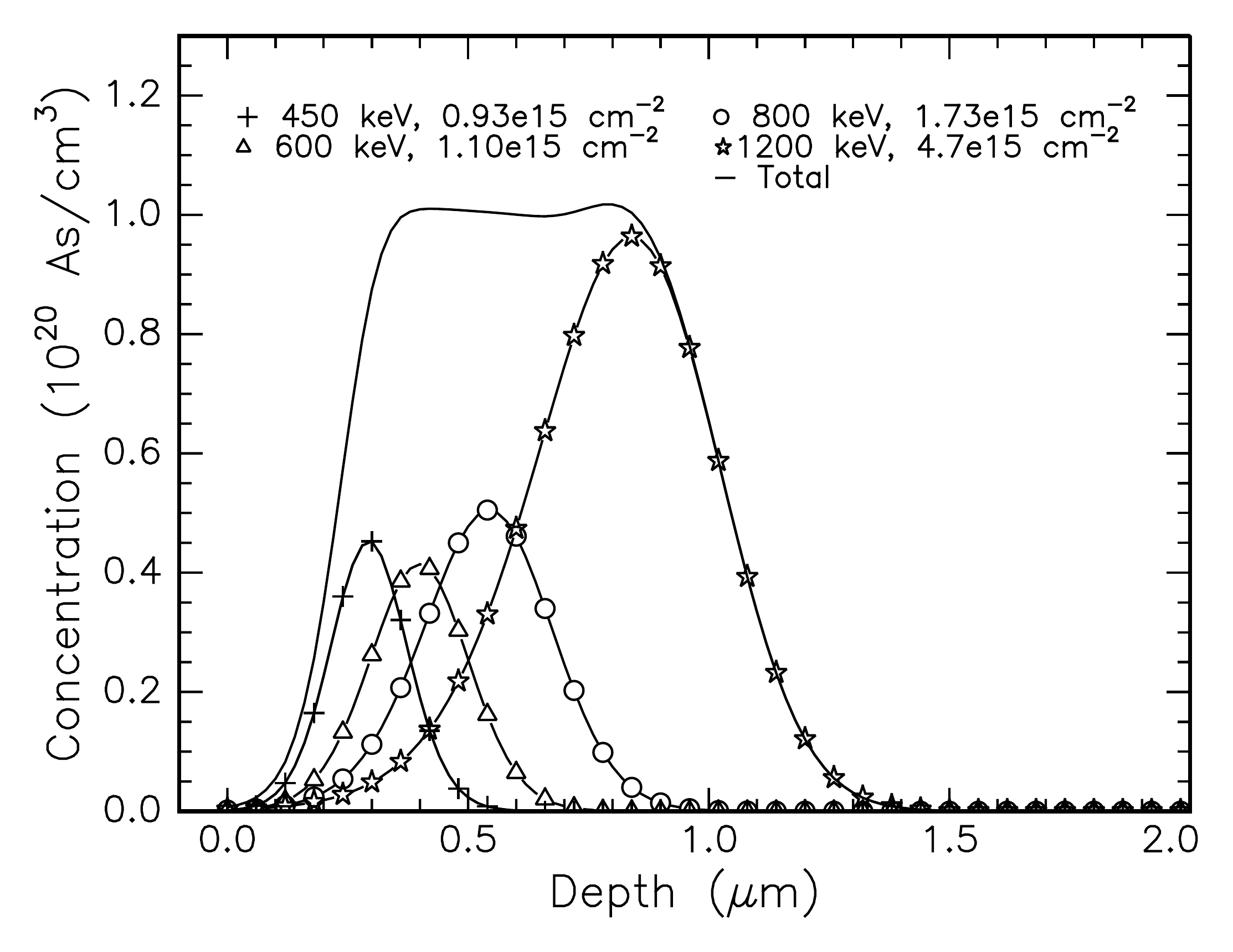}}
\end{center}
\caption[As constant concentration profile]{Theoretical calculation of
the four implants used to create a constant As concentration profile
over the depth range 0.45-0.85~$\rm \mu m$ as used by McCallum in
Ref.~\cite{NIMB:mccallum}.}
\label{figs/profileas.ps}
\end{figure}

\begin{figure}[h!]
\begin{center}
\rotatebox{90}{\includegraphics[height= 17cm]{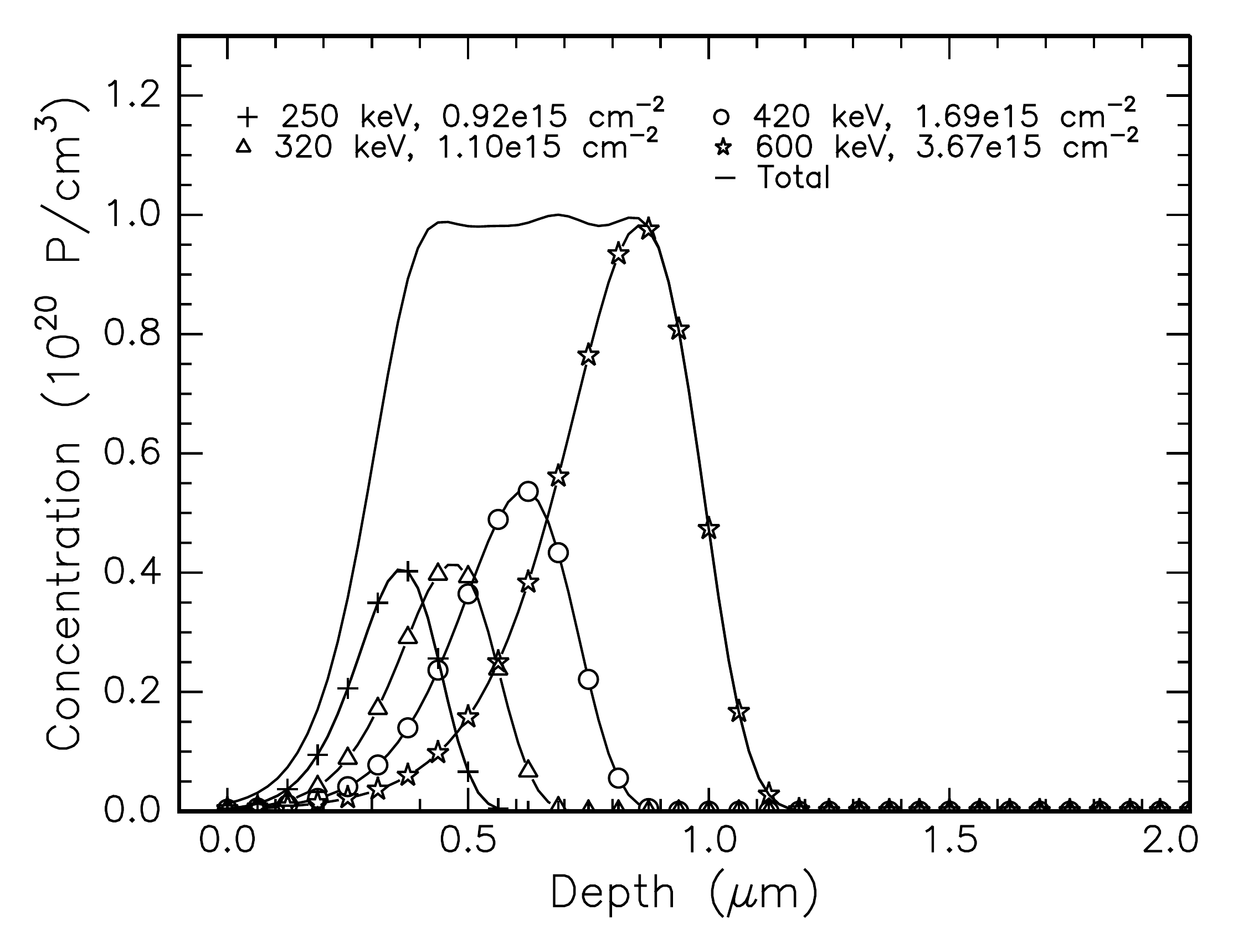}}
\end{center}
\caption[P constant concentration profile]{Theoretical calculation of
the four implants used to create a constant P concentration profile
over the depth range 0.5-0.8~$\rm \mu m$.}
\label{figs/profilep.ps}
\end{figure}

\begin{figure}[!t]
\begin{center}
\rotatebox{90}{\includegraphics[height=17cm]{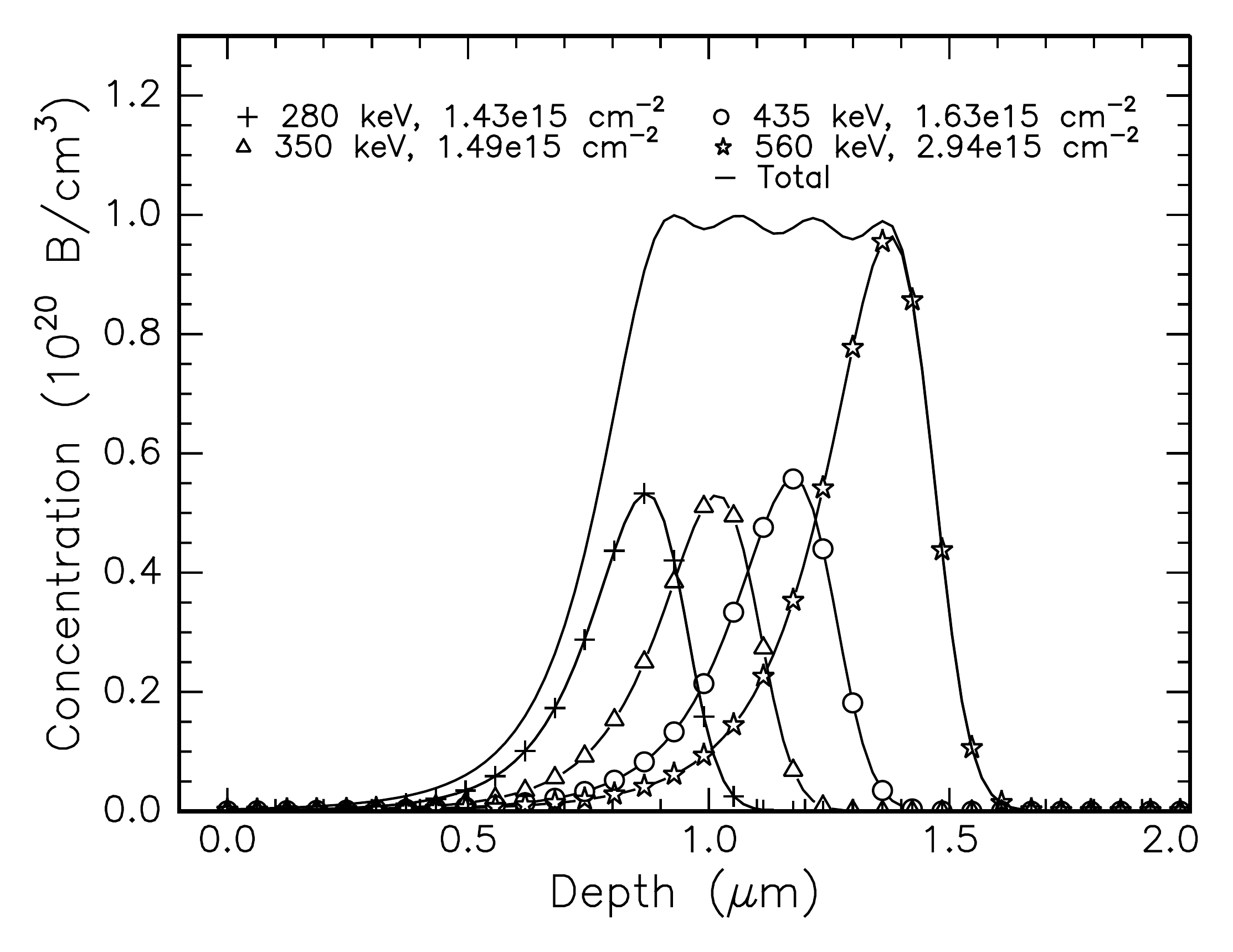}}
\end{center}
\caption[B constant concentration profile]{Theoretical calculation of
the four implants used to create a constant B concentration profile
over the depth range 0.95-1.35~$\rm \mu m$.}
\label{figs/profileb.ps}
\end{figure}

\begin{figure}[h!]
\begin{center}
\rotatebox{90}{\includegraphics[height= 17cm]{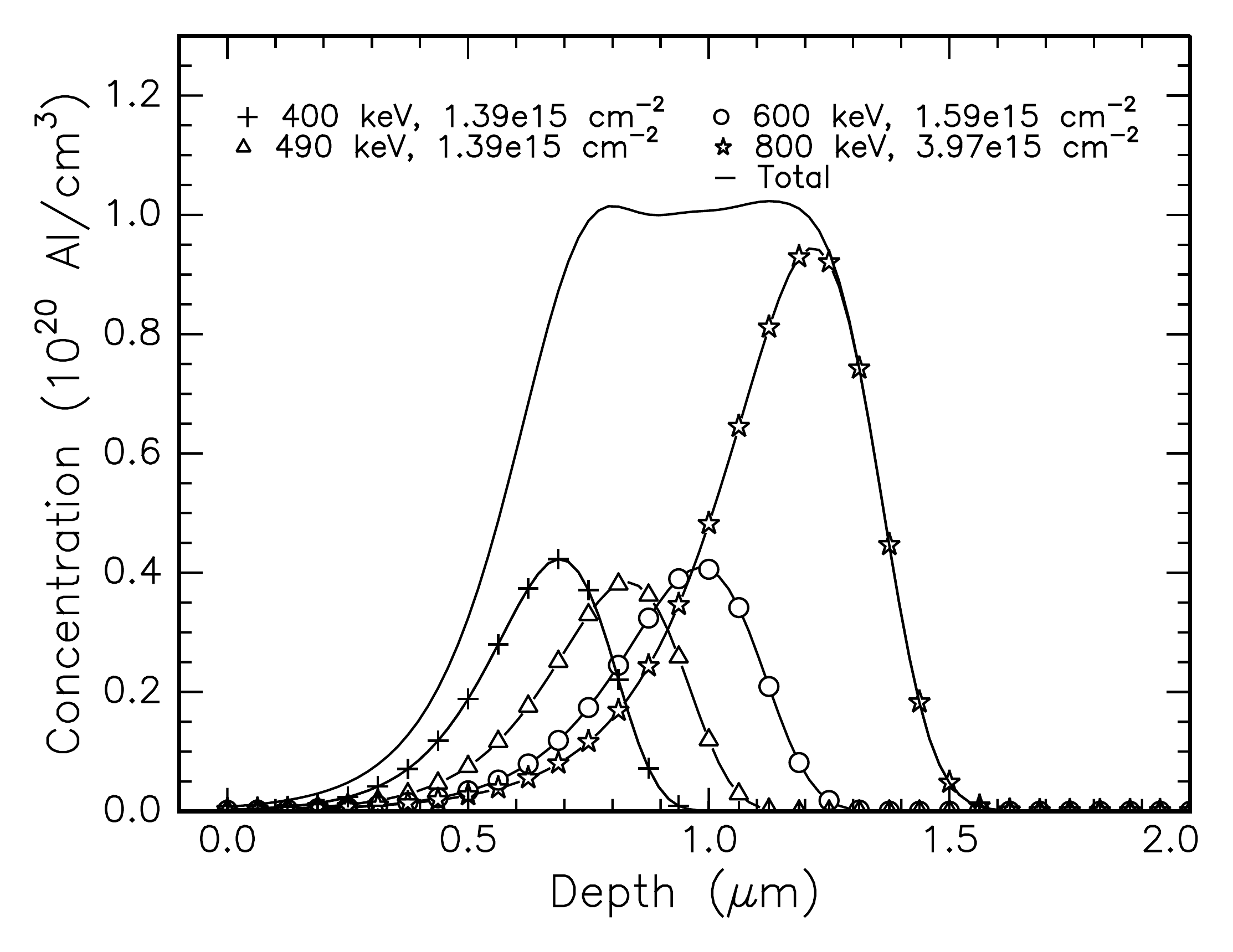}}
\end{center}
\caption[Al constant concentration profile]{Theoretical calculation of
the four implants used to create a constant Al concentration profile
over the depth range 0.8-1.15~$\rm \mu m$.}
\label{figs/profileal.ps}
\end{figure}

To fabricate the samples, surface $a$-Si layers were produced by
forming intrinsic amorphous layers $\rm 2.2\: \mu m$ thick using the
implantation schedule: $\rm Si^{28}(500\: keV,\: 2\times 10^{15}\:
cm^{-2},\: -195^{o}C)$ and $\rm Si^{28}(2\: MeV,\: 3\times 10^{15}\:
cm^{-2},\: -195^{o}C)$. Constant concentration profiles of P, B or Al
were then created by multiple energy implantation. The expected
concentration profiles calculated using the Profile code\cite{profile}
are shown in Figs.~\ref{figs/profileas.ps}-\ref{figs/profileal.ps} for
As, P, B and Al, respectively. The implant schedule used to create a
constant concentration profile of $\rm 1 \times 10^{20} \; cm^{-3}$
over a certain depth is indicated at the top of each figure. The
schedule used for As implants in Ref.~\cite{NIMB:mccallum} is shown
for completeness. The implant fluences were scaled to obtain a range of
peak concentrations between $\rm 1 \times 10^{19}$ and $\rm 3 \times
10^{20} \; cm^{-3}$. SPE rate enhancement is generally immeasurable at lower concentrations and approaches the solid solubility limit at concentrations higher than this range. The dopant concentrations were relatively constant over the depth range:
0.45-0.85, 0.5-0.8, 0.95-1.35 and 0.8-1.15~$\rm \mu m$ for As, P, B
and Al, respectively. Preliminary secondary ion mass spectrometry
(SIMS) data from P and B implanted samples compare well with the shape
and depth range of these curves although a standard to calibrate the
concentration axis was not obtained. SIMS on the Al doped layers was
not performed. For As, RBS-C was used to compare expected and actual
profiles and agreement was obtained to within $\sim$10\%.

After implantation samples were annealed $in-situ$ at $\rm 600^{o}$C
for 1~hour under UHV conditions in order to completely crystallize the
layer. Buried amorphous layers were then formed by using the
implantation schedule: $\rm Si^{28}(600\: keV,\: 5\times 10^{14}\:
cm^{-2},\: -10^{o}C)$ and $\rm Si^{28}(2\: MeV,\: 3\times 10^{15}\:
cm^{-2},\: -10^{o}C)$ using a flux of $\rm 1\: \mu A.cm^{-2}$. This
schedule produced a buried amorphous layer 0.87~$\rm \mu m$ thick with
a 0.3~$\rm \mu m$ thick $c$-Si capping layer.

The SPE rates were determined in air using a time-resolved
reflectivity (TRR) system equipped with two lasers collecting data
simultaneously at $\rm \lambda=1152\; nm$ and $\rm \lambda=632.8\; nm$.\cite{APL:mccallum} The samples were held on a resistively heated vacuum chuck while anneals were performed over a temperature range of $\rm 460-660^{o}C$ in $\rm 20^{o}C$ increments. The temperature of the samples during the anneals was calibrated by comparing the reading of a type-K thermocouple embedded in the sample stage with the melting points of various
suitably encapsulated metal films evaporated onto Si wafers. The error
associated with the temperature reading was found to be $\rm \pm
1^{o}$C.

\section{Results and Discussion}
\label{results}
The kinetics of dopant-enhanced SPE were measured over the temperature
range 460-660$\rm ^{o}$C for samples containing a number of different
P, B or Al concentrations in the range $\rm 1-30 \times 10^{19} \;
cm^{-3}$. These were compared to the As-enhanced SPE measurements
presented in Ref.~\cite{NIMB:mccallum} by McCallum collected on the
same TRR system.

\begin{figure}[!h]
\begin{center}
\rotatebox{180}{\includegraphics[height= 15cm]{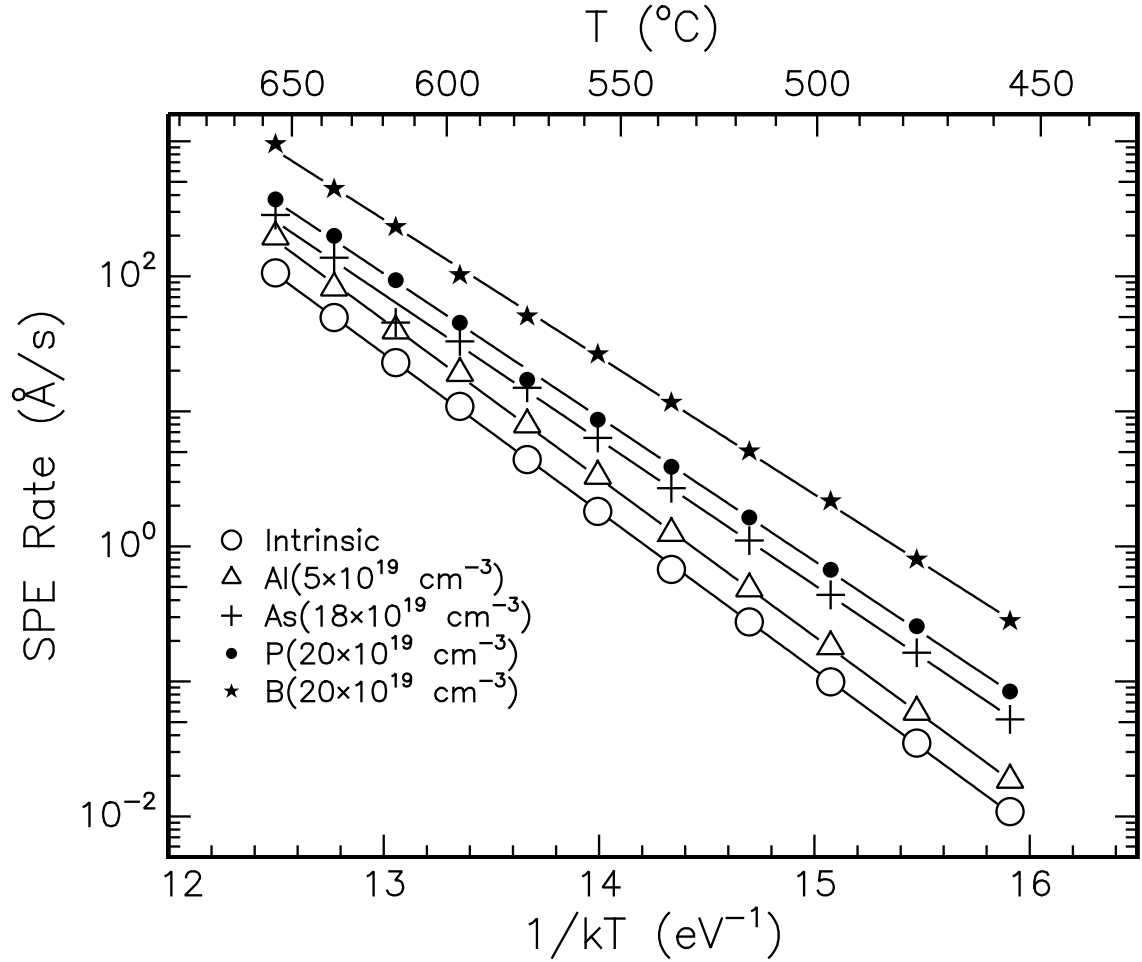}}
\end{center}
\caption[Arrhenius plot of doped buried $a$-Si layers]{Arrhenius plot
showing the temperature dependence of the SPE regrowth rate for
intrinsic and and selected doped buried $a$-Si layers. The solid lines
are least-square fits of the data using an Arrhenius type expression.}
\label{figs/arrhenius}
\end{figure}

The temperature dependence of the SPE regrowth rate for intrinsic and
doped buried $a$-Si layers is shown in Fig.~\ref{figs/arrhenius}. For
clarity, only one concentration for each dopant studied is
plotted. Fitting an Arrhenius type equation of the form $v=v_{o}
\exp(-E_{a}/kT)$ to the intrinsic SPE data yielded a velocity
prefactor of $v_{o}=(4 \pm 1) \times 10^{16}$~$\rm \AA$/s and an
activation energy of $E_{a}=(2.68 \pm 0.04)$~eV. The errors associated
with these values were calculated by considering the $\pm$1\%
temperature reproducibility, the RMS noise in the determined velocity
curve and the errors associated with the fitting procedure. These
values compare quite well to those reported by Roth {\it et al.} for
thick surface $a$-Si layers which were $4.64\times 10^{16}$~\AA/s and
2.7~eV.\cite{APL:roth}

The greatest $c$-$a$ interface velocity enhancements were found for
samples implanted with B. It was also found that Al concentrations
greater than $\rm 5 \times 10^{19} \; cm^{-3}$ caused the TRR signal
to collapse. This indicated that the $c$-$a$ interface had become
rough. SPE rate retardation and interface segregation has
previously been observed for Al concentrations above $\rm \sim 2
\times 10^{20} \; cm^{-3}$.\cite{review:roth}

\begin{figure}[!t]
\begin{center}
\rotatebox{180}{\includegraphics[height= 15cm]{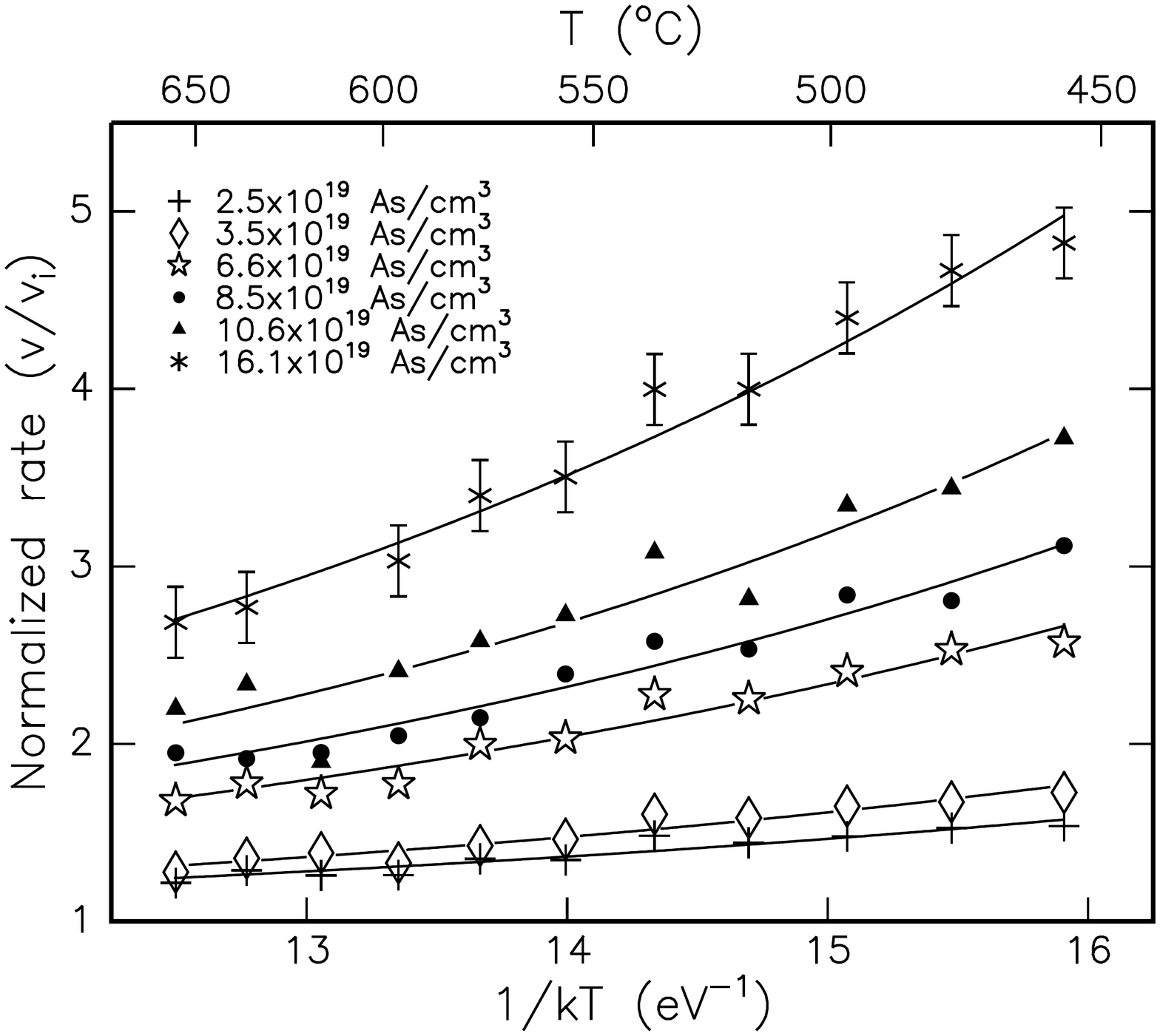}}
\end{center}
\caption[Normalized As-enhanced SPE rates]{As-enhanced SPE rates for
the front interfaces of buried $a$-Si layers normalized to the
corresponding intrinsic SPE rate from Ref.~\cite{NIMB:mccallum}.}
\label{asdata}
\end{figure}

\begin{figure}[!h]
\begin{center}
\rotatebox{180}{\includegraphics[height= 15cm]{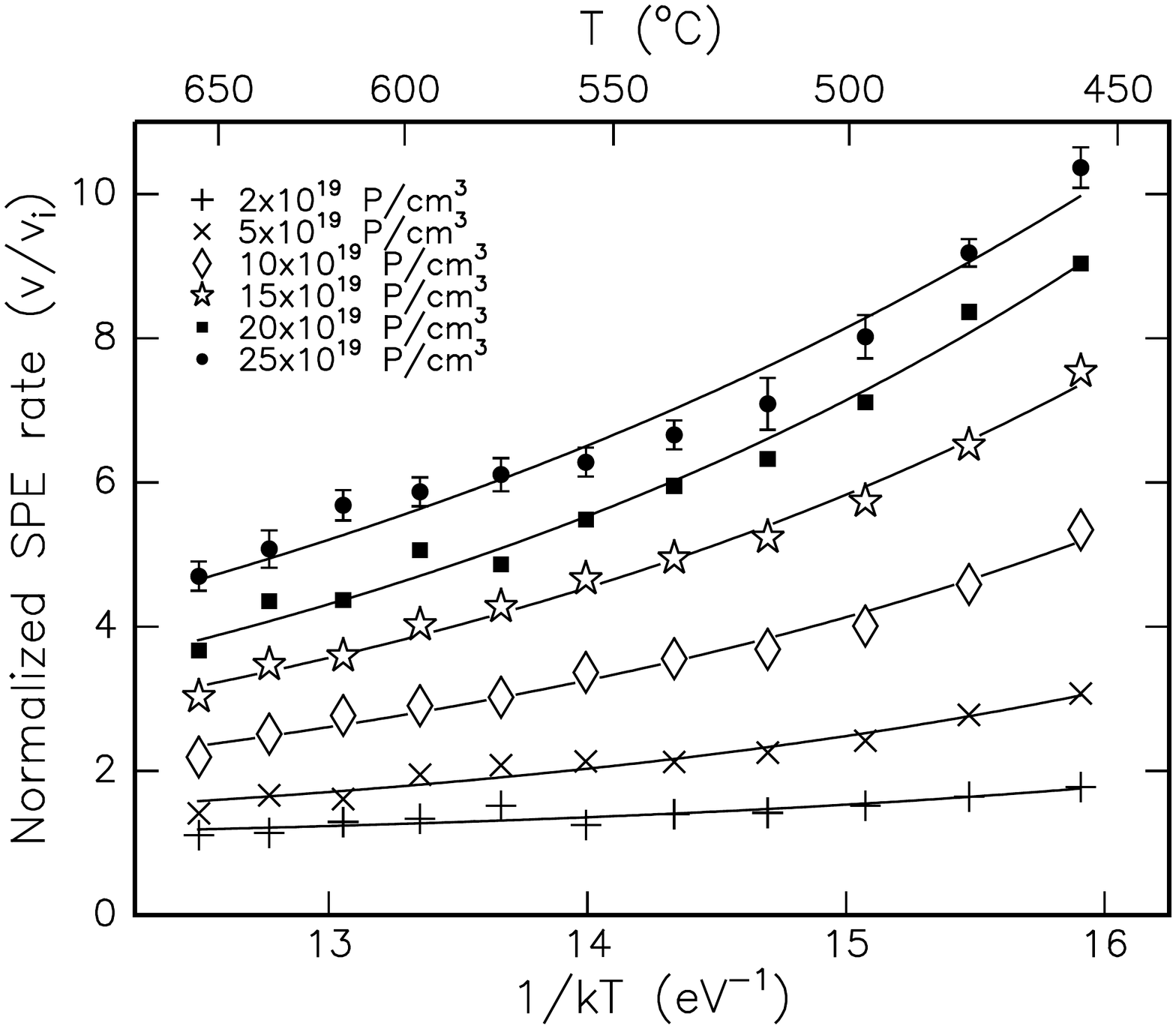}}
\end{center}
\caption[Normalized P-enhanced SPE rates]{Phosphorus-enhanced SPE
rates for the front interfaces of buried $a$-Si layers normalized to
the corresponding intrinsic SPE rate.}
\label{pdata}
\end{figure}

\begin{figure}[!t]
\begin{center}
\rotatebox{180}{\includegraphics[height= 15cm]{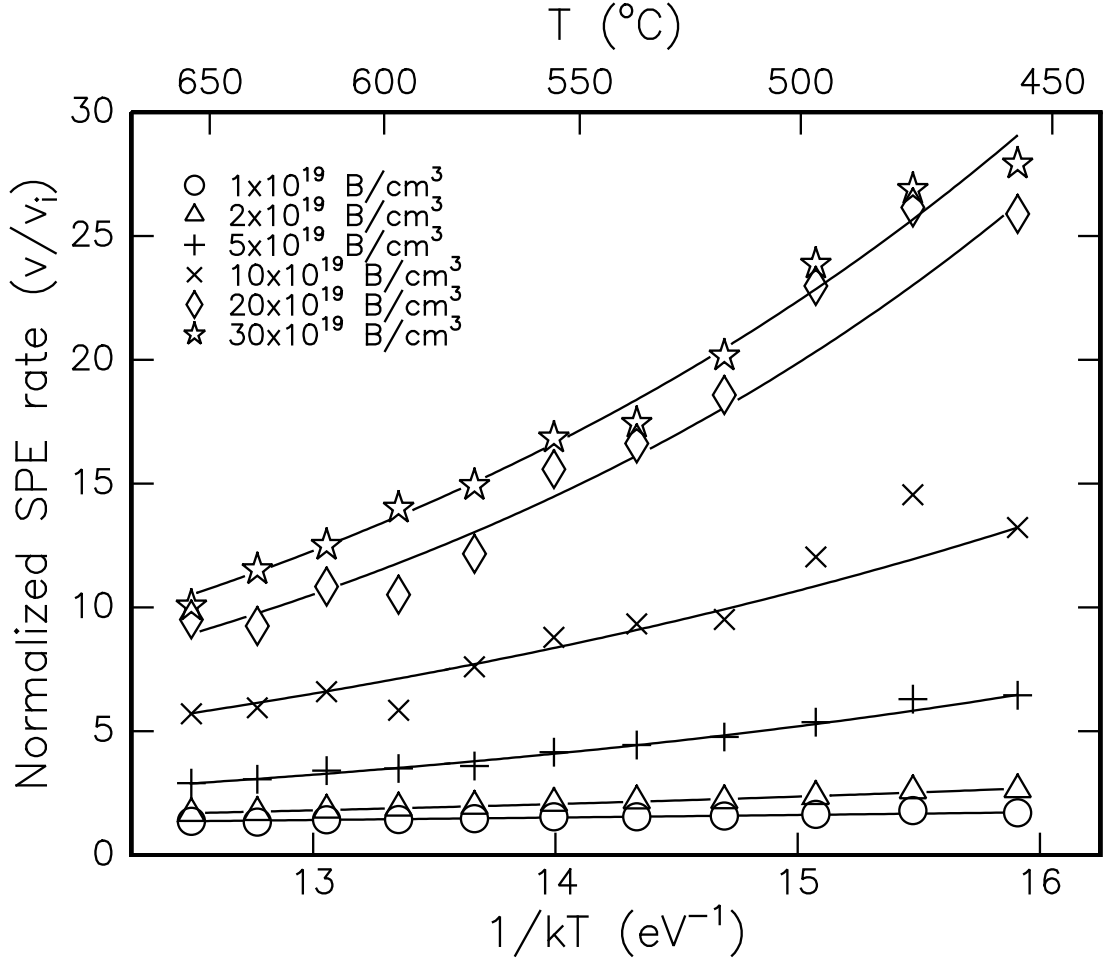}}
\end{center}
\caption[Normalized B-enhanced SPE rates]{B-enhanced SPE rates for the
front interfaces of buried $a$-Si layers normalized to the
corresponding intrinsic SPE rate. Due to the scale of the y-axis,
errors are about the size of the symbols.}
\label{bdata}
\end{figure}

\begin{figure}[!h]
\begin{center}
\rotatebox{180}{\includegraphics[height= 15cm]{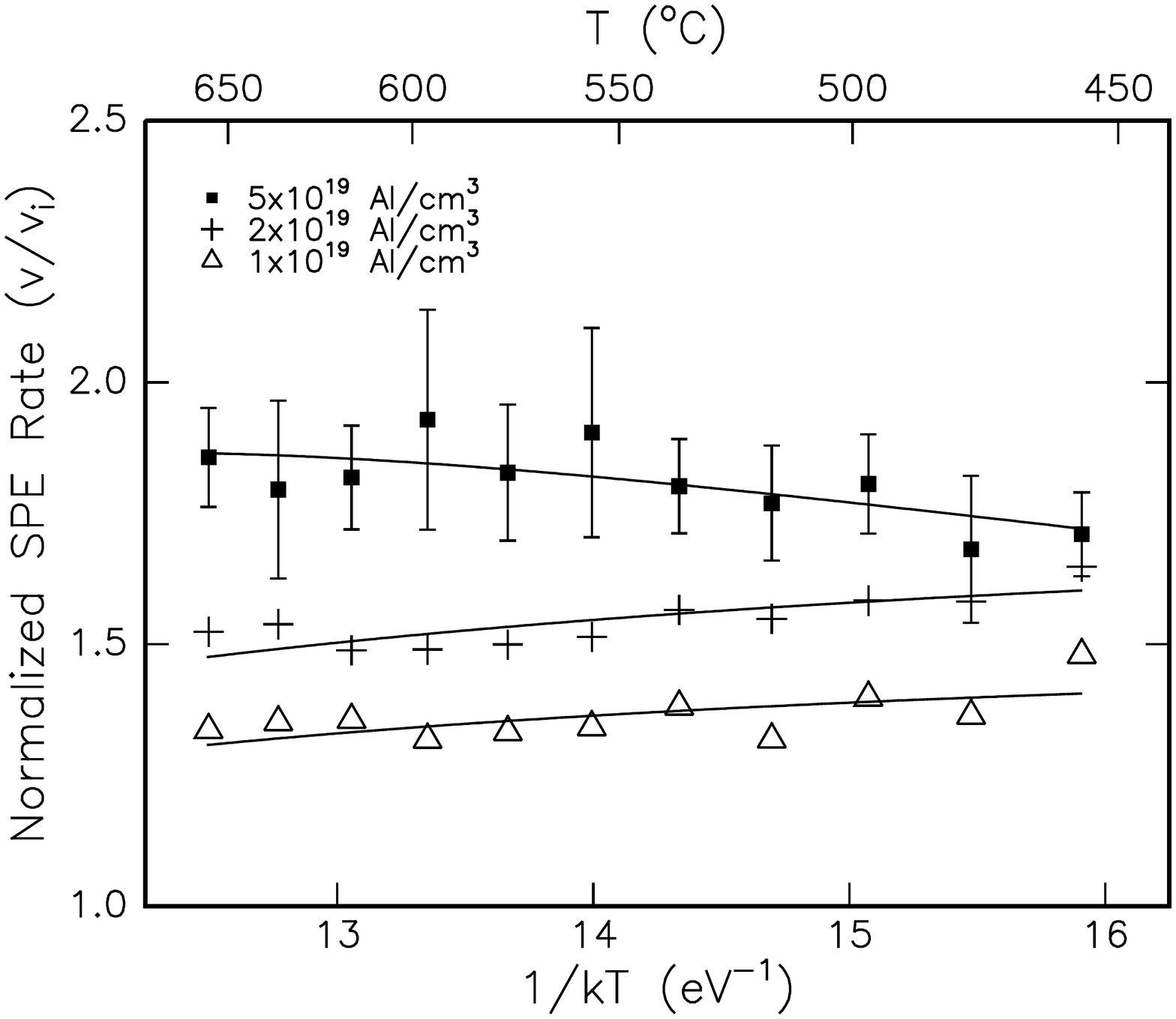}}
\end{center}
\caption[Normalized Al-enhanced SPE rates]{Al-enhanced SPE rates
for the front interfaces of buried $a$-Si layers normalized to the
corresponding intrinsic SPE rate.}
\label{aldata}
\end{figure}

Figures~\ref{asdata},~\ref{pdata},~\ref{bdata} and~\ref{aldata} show
the dopant-enhanced SPE rates for buried $a$-Si layers doped with
constant concentrations of As, P, B and Al, respectively.  Errors for
the As data from Ref.~\cite{NIMB:mccallum} are estimated by
considering the reproducibility of the data. Errors for the P, B and
Al also take into account the RMS noise in the determined velocity
curve. For clarity, errors are presented for only one
concentration. Rates were normalized to the intrinsic SPE rate
values. The As, P and B enhanced SPE data exhibit the typical trends
with the greatest enhancement occurring for the lowest temperatures
and highest concentrations. The Al-enhanced SPE rate at a
concentration of $\rm 5 \times 10^{19} \; cm^{-3}$, on the other hand,
shows the greatest enhancement at the highest temperature studied
although the temperature dependence is not significant. This is
unexpected as the greatest Fermi level shifts, and therefore the
greatest rate enhancements, occur for the lowest temperatures. SPE
enhancement at lower Al concentrations is consistent with other
dopants, however, the general variation with temperature is
different. We attribute this anomalous behaviour to the fact that
interface roughening occurs during the SPE regrowth of Al doped $a$-Si
layers.\cite{review:roth} Al is atypical of other dopants we have
examined and clearly there are other factors that are influencing the
SPE growth process in the case of this dopant.
The greatest SPE rate enhancement is observed for samples implanted
with B which, at 460 $\rm ^{o}$C and a concentration of $\rm 30 \times
10^{19} \; cm^{-3}$, is about 30 times greater than the intrinsic
value.

The solid lines in Figs.~\ref{asdata}~-~\ref{aldata} are fits using the
GFLS model presented in section~\ref{gfls} incorporating the degenerate semiconductor statistics discussed in the appendix. The weighted averages of the energy level and degeneracy values obtained from these fits are
presented in Table~\ref{gandden} for $n$-type dopants and
Table~\ref{ganddep} for $p$-type dopants. In the first instance, the
energy level and degeneracy of the SPE defect were both allowed to
vary in the fitting routine. These fits are plotted in
Figs~\ref{asdata}~-~\ref{aldata} with the dopant-enhanced SPE
data. All fits to our data yielded reasonable values although the
energy level extracted from fits to Al-enhanced SPE data were found to
be below the top of the valence band. This value may represent a
combination of the SPE defect and effects responsible for the
anomalous Al-enhanced SPE data as mentioned earlier.

\begin{table}[!t]
\caption[Degeneracy and energy level for the negative charge state of
the SPE defect]{\label{gandden} The weighted averages of the
degeneracy, $g$, and energy level, $E^{-}$, of the defect identified
by the GFLS model from fits to the $n$-type dopant-enhanced SPE
data. Energies are in eV and are referenced to the edge of the
conduction band.}
\begin{tabular}{rcc|c|cc}
\hline \hline
   &$E_{c}-E^{-}$ (eV), &  $g$            & $E_{c}-E^{-}$ (eV)  &$E_{c}-E^{-}$ (eV) &\\
   &                    &                 &(g=0.4)              &(g=0.5)            &\\
\hline
As data & $0.16\pm0.01$      & 0.53$\pm$0.07   &$0.18\pm0.01$        &$0.17\pm0.01$      &\\
P data & $0.23\pm0.02$      & 0.25$\pm$0.06   &$0.22\pm0.01$        &$0.21\pm0.01$      &\\
\hline \hline
\end{tabular}
\end{table}

\begin{table}[!t]
\caption[Degeneracy and energy level for the positive charge state of
the SPE defect]{\label{ganddep} The weighted averages of the
degeneracy, $g$, and energy level, $E^{+}$, of the defect
identified by the GFLS model from fits to the $p$-type dopant-enhanced
SPE data. Energies are in eV and are referenced to the edge of the
valence band.}
\begin{tabular}{rcc|c|cc}
\hline\hline
&$E^{+}-E_{v}$ (eV),&$g$&$E^{+}-E_{v}$ (eV)&$E^{+}-E_{v}$ (eV)&\\
&                  &   &(g=1.5)           &(g=1)             &\\
\hline
B data&$0.17\pm0.01$&1.5$\pm$0.2 &$0.20\pm0.02$&$0.23\pm0.02$&\\
Al data&$-0.08\pm0.01$  & 16$\pm$3 &$0.12\pm0.01$  &$0.14\pm0.01$  &\\
\hline \hline
\end{tabular}
\end{table}

The errors associated with the values in Tables~\ref{gandden}
and~\ref{ganddep} took into account the $\rm \pm 1^{o}$C temperature
reproducibility and the 3\% variation between the use of the
semiconductor parameters of Green and those of Alex and Green as discussed in the appendix. A
calculation to see how the free fitting parameters might respond to a
10\% dopant concentration error was also made. The greatest variations
in $E^{\pm}$ and $g$ were found in the low fluence regime where the Fermi
level shifts are more sensitive to changes in dopant concentration. At
higher fluences the Fermi level asymptotes to the band edges. Generally,
for variations of 10\% in the dopant concentration the defect level
was found to shift by 0.015~eV to 0.001~eV for fluences between 2 and 30 $\rm \times 10^{19} \; cm^{-3}$, respectively. Conversely, the degeneracy
changed by 0.01 and 0.1 for fluences in the same range. These errors were
also included in the values presented in Tables~\ref{gandden}
and~\ref{ganddep}.

The values of $E^{\pm}$ and $g$ in Tables~\ref{gandden}
and~\ref{ganddep} are reasonable in that the degeneracies are not
expected to be large and the values of $E^{\pm}$ are consistent with
the energy levels of typical charged defects in $c$-Si such as
V$_{2}^{2-}$ (which, for example is
($E_{c}-0.22$~eV)).\cite{JAP:huppi} The degeneracy for As is close to
0.5 which is consistent with a negatively charged DB defect. The
degeneracy for B is 1.5 which is somewhat higher than the value of
unity that we expect of a positively charged DB. But, given the number
of factors involved in arriving at these fitted values the agreement
with expected ranges of values is remarkable.

There is a slight discrepancy between $E^{-}$ and $g$ determined from
As-enhanced SPE data with values reported by McCallum in
Ref.~\cite{NIMB:mccallum}. This is mainly a result of including the
effective mass temperature dependence and the concentration dependence
of the BGN in the calculations presented here.

According to the GFLS model the $E^{-}$ and $g$ values for dopants of
the same type should be equal. The energy levels determined from As and P
data show a discrepancy even after all the relevant errors are taken into
account. However, if $g$ is held fixed during fitting they become more
consistent as can be seen in the middle column of the tables. This discrepancy suggests that some appropriate parameters
or temperature dependences may not be properly incorporated into the model.

The degeneracy was also set to the values expected of a DB. Again,
fits to the As, P and B data sets with degeneracy values fixed at
these values were reasonable while the energy level values tended to
increase with decreases in the degeneracy factor. For the DB
degeneracy values, the energy level of the positively charged defect
tended to be greater than its negatively charged counterpart. This
trend is similar to that predicted by Mosley and Paesler in their
electric field model except that their energy levels were much closer
to the center of the band gap.\cite{JAP:mosley} Fits to Al-enhanced
data with fixed degeneracy values were quite poor as the lower
degeneracy forced the greatest rate enhancement to be at the lowest
temperatures - the opposite trend to the actual data.

\begin{figure}[!ht]
\begin{center}
\rotatebox{90}{\includegraphics[height=10cm]{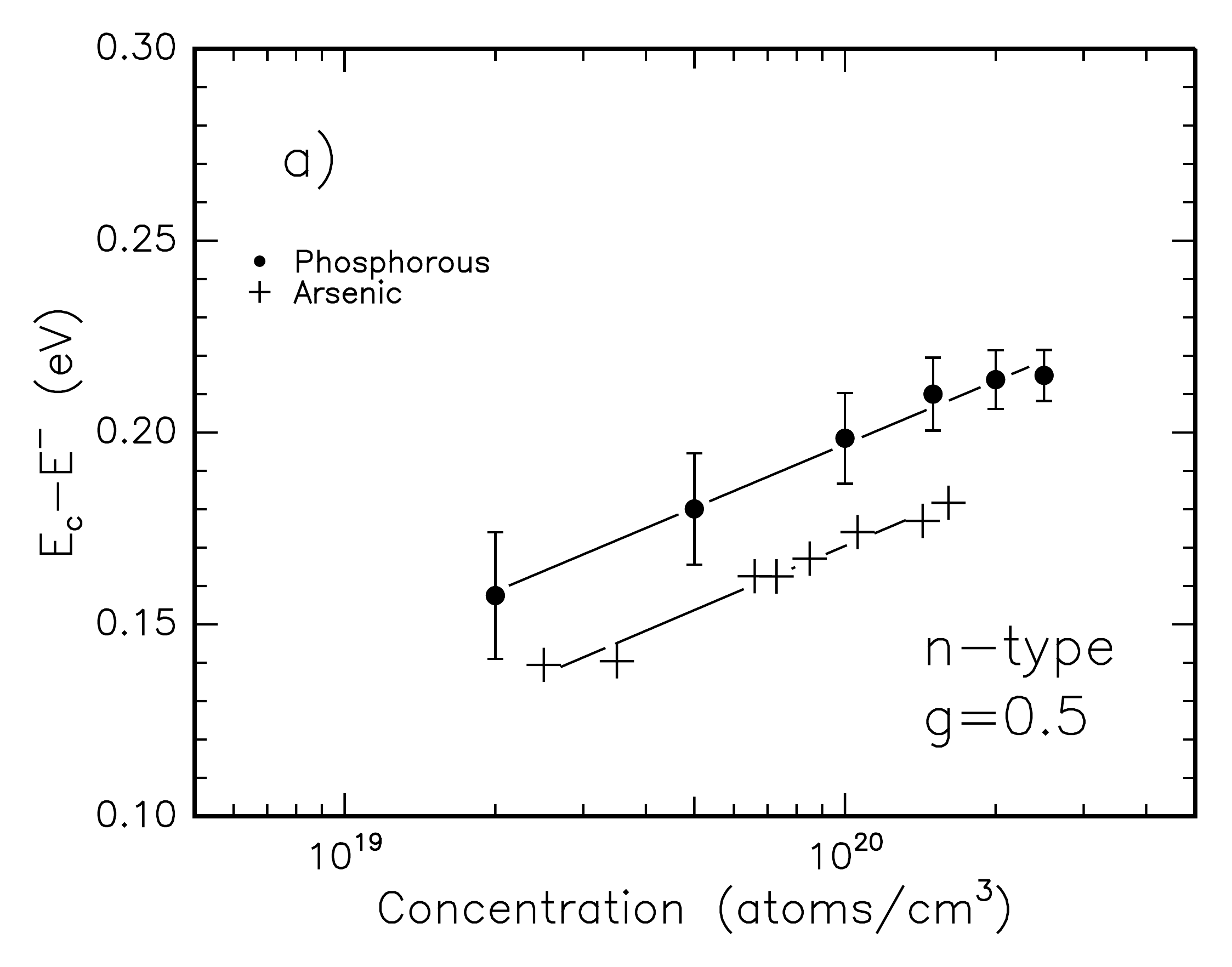}}\\
\rotatebox{90}{\includegraphics[height=10cm]{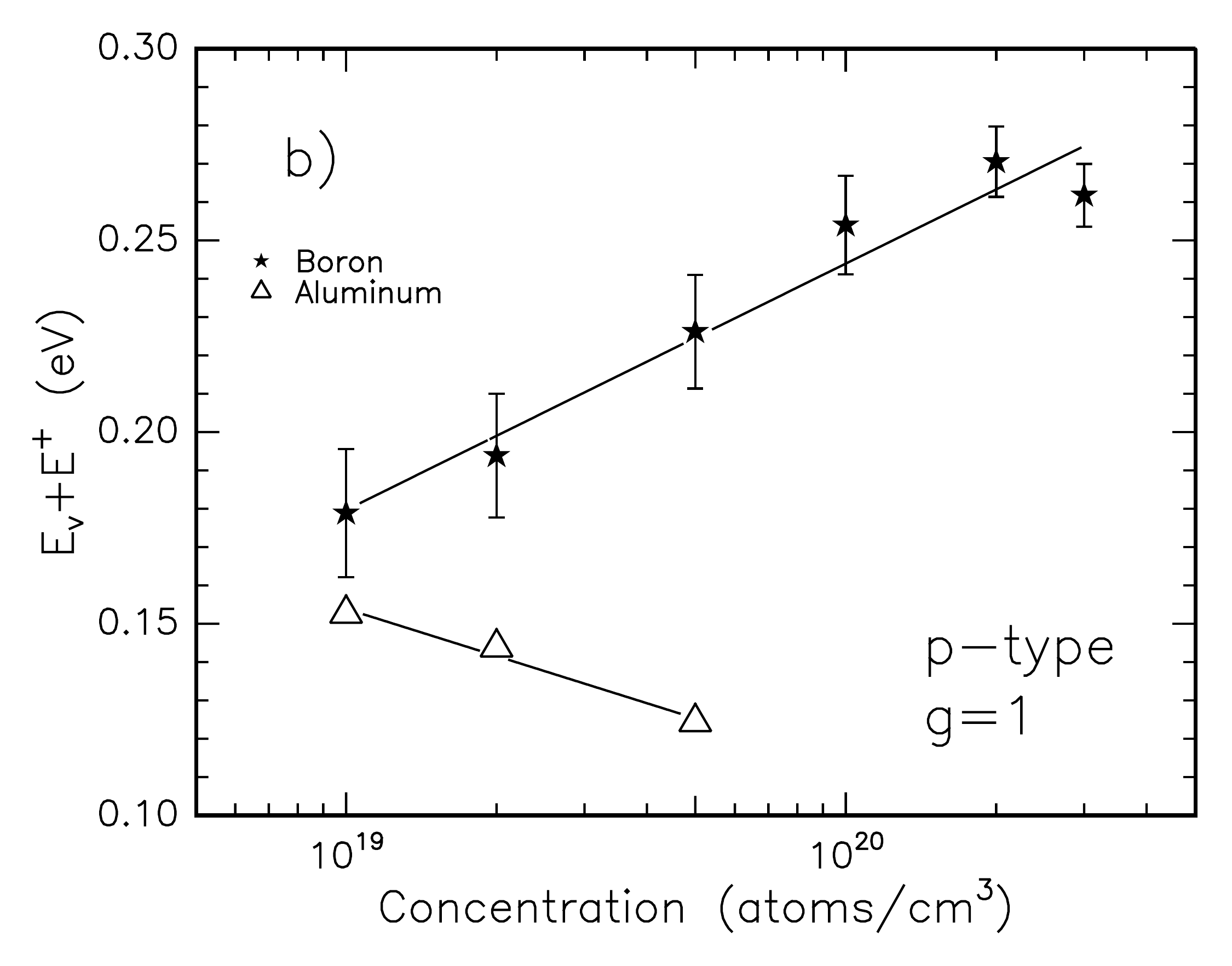}}
\end{center}
\caption[Defect energy level as a function of dopant concentration]{a)
The defect energy level identified by the GFLS model with the
degeneracy fixed at g=0.5 for phosphorus ($\bullet$) and
arsenic (+) and b) at g=1 for boron
($\star$) and aluminum ($\bigtriangleup$) as a function of dopant
concentration. Solid lines are a guide only.}
\label{figs/pndevsconc}
\end{figure}

Figure~\ref{figs/pndevsconc} shows the SPE defect level as a function
of dopant concentration extracted from fitting Eq.~\ref{fit} to the
data with a fixed value of the degeneracy. This figure illustrates the
systematic error that exists in fitting our data. Apart from the Al
data, which shows anomalous SPE behaviour, all other trends show a
similar gradient. In terms of the dopant concentration, one way of
removing these trends from the data would be by modifying the values
of $N_{d}$ used in our calculations. To fit the trends we would
require $N_{d}$ to be underestimated and for the degree of underestimation to increase with dopant concentration. For example, to make $E^{-}=0.14$
for $N_{d} = 16.1 \times 10^{19}$ $\rm As/cm^{3}$ the concentration
must be underestimated by a factor of 2.8. This suggests that the
appropriate dopant concentration dependences may not be included in
the physical parameters outlined in the appendix. In
addition, the concentration dependence of the BGN shifts $E^{-}$ by a
value too small to explain the observed effect. Band bending may also
cause such an effect as discussed below. Further, the concentration dependence of the effective mass is not expected to play a major role below concentrations of $\rm 1 \times 10^{21} \; cm^{-3}$ (see the Appendix).

Figure~\ref{band} shows the energy levels of the defect responsible for
the SPE process according to the GFLS model using a fixed degeneracy value
expected of a dangling bond-type defect (0.5 and 1 for the negative and
positive defect, respectively). The area between the $E^{-}_{As}$ and
$E^{-}_{P}$ is shaded to indicate that we do not expect these levels to
be different and therefore $E^{-}$ may lie somewhere in this range. To
provide some reference point for these energy levels with respect to known defect levels in $c$-Si the energy levels of some vacancy-related
defects are also shown. These energy levels were measured in $n$-type
silicon using deep-level transient spectroscopy and their associated
energy levels are: V-related ($E_{c}-0.35$~eV),\cite{JAP:palmetshofer}
V-related ($E_{c}-0.19$~eV), V$_{2}^{2-}$ ($E_{c}-0.22$~eV), V$_{2}^{-}$
($E_{c}-0.42$~eV) and V$_{2}$-related ($E_{c}-0.47$~eV).\cite{JAP:huppi}
The two vacancy related defects giving energy levels at $E_{c}-0.19$~eV
and $E_{c}-0.22$~eV are very close to the levels of the negatively charged
defect predicted by the GFLS model. Vacancy related defect levels are
included here just to show that values for $E^{-}$ are consistent with
band gap positions of some known defects. We do not know of the existence
of any energy level values for DBs in $c$-Si in the literature.
Recently, McCallum mentioned that the GFLS model could be extended to
predict the energy level of the SPE defect at the $c$-$a$ interface by
accounting for band bending between the amorphous and crystalline
phases.\cite{NIMB:mccallum} A $c$-$a$ interface band structure has
been proposed by Williams and Elliman which is similar to a $p-n$
junction with the Fermi level on one side of the interface taking 
the $a$-Si value.\cite{PRL:williams} In intrinsic material the Fermi
level on both the crystalline and amorphous side of the interface are
close to mid-gap. When doped the Fermi level on the $c$-Si side will
shift in order to satisfy charge neutrality. In $a$-Si, it will
generally remain pinned to mid-gap due to the high density of
localized states in the center of the band gap. The band structure at
the position of the SPE defect residing at the interface can be varied
with a weighting function of the form

\begin{equation}
\label{amorphweight}
E_{F}=W(E_{f}-E_{fa})+E_{fa}
\end{equation} 

\noindent where $W$ is the weighting factor and $E_{fa}$ is the Fermi
level of $a$-Si. A value of $W=1$ would result in the Fermi level at
the interface being equal to the bulk crystalline value. This is
equivalent to the original GFLS model. A value of $W<1$ would result
in the value at the interface being shared by the crystalline and
amorphous phases. Finally, with a value of $W=0$, the Fermi level is
pinned to mid-gap in the amorphous phase and doping may have little
effect.

Details of the band structure of $a$-Si are relatively sparse and
lacking in consistency. For hydrogen-free as-implanted intrinsic
$a$-Si Stolk {\em et al.} have determined the value of the band gap
at room temperature to be $E_{g}=1.2$~eV.\cite{JAP:stolk} An empirical
formula for thermal BGN in hydrogenated $a$-Si has been reported by
Bube {\em et al.}\cite{JAP:bube} They estimate a decrease of about 50~meV in the
band gap between temperatures of 300 and 400~K.
Likewise, Premachandran {\em et al.} have reported the temperature
dependence of the mobility edge of $a$-Si to be $dE_{c}/dT=8\times
10^{-4}$~eV/K and that both $E_{c}$ and $E_{v}$ contribute equally to
the narrowing.\cite{PRB:premachandran} The Fermi level shift was also
found to be $dE_{f}/dT=3\times 10^{-4}$~eV/K. There is also some
evidence that the effective mass is larger in $a$-Si than it is in
$c$-Si which would affect the effective DOS values.\cite{APL:chen}
Data on doping effects in H-free $a$-Si in the concentration range
relevant to SPE are extremely sparse. In short, a complete and
reliable picture of the amorphous band structure is lacking from the
literature. Once known, the inclusion of Eq.~\ref{amorphweight} into
the GFLS model calculations would be straight forward and may provide
a more complete picture of the SPE process.

\begin{figure}
\begin{center}
\rotatebox{0}{\includegraphics[height= 16cm]{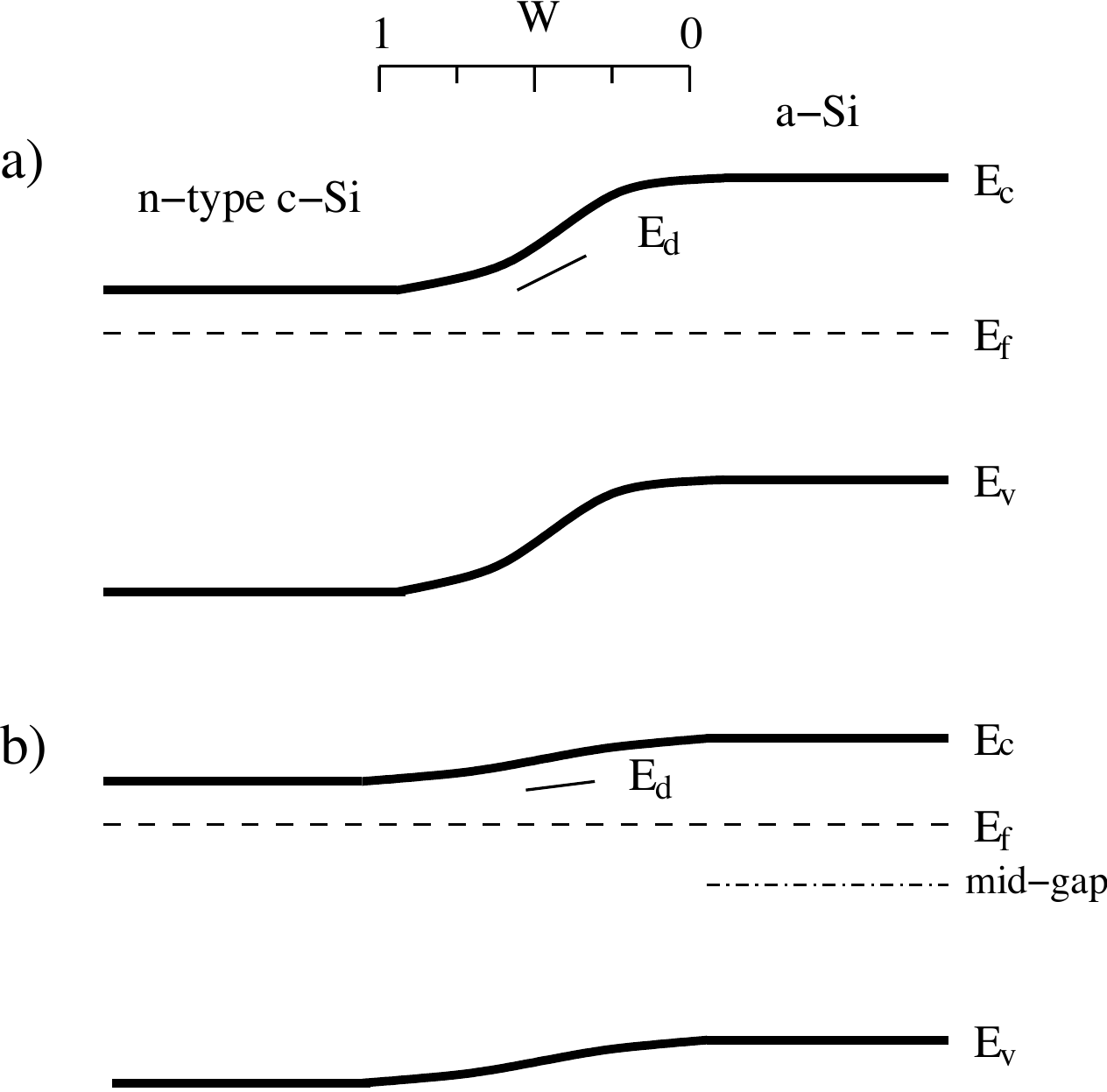}}
\end{center}
\caption[Band gap structure at the $c$-$a$ interface]{a) Proposed band
gap structure at the $c$-$a$ interface for $n$-type Si. The weighting
factor, $W$, in Eq.~\ref{amorphweight} is indicated. The Fermi level
remains constant across the interface and the SPE defect level is also
shown in the middle of the interface region ($\rm W=0.5$). b) shows
the same interface with the Fermi level unpinned from mid gap on the
amorphous side of the interface. The Fermi level moves closer to the
SPE defect level.}
\label{bandbending}
\end{figure}

Figure~\ref{bandbending}a) shows a possible band structure diagram at
the $c$-$a$ interface for $n$-type Si. The Fermi level remains
constant across the interface region causing band bending to
occur. The weighting factor scale is indicated above this band
bending. The SPE defect level is placed at
$W=0.5$. Fig.~\ref{bandbending}b) shows the possible band structure of
the same material if the Fermi level were to come unpinned on the
amorphous side of the interface through the filling of mid-gap states
as found for high dopant concentrations ($>$1~at.\%) by Coffa and
co-workers.\cite{NIMB:coffa,APL:coffa} H is also known to passivate
defects in $a$-Si and could also modify the band bending at the
interface. In fact, the Fermi level has been shown to have a linear
dependence on the defect density in $a$-Si:H which can be controlled
by the H concentration or the substrate temperature, at least in
deposited $a$-Si:H films.\cite{JAP:bube}
If the SPE defect resides within the band bending region and unpinning
occurs the Fermi level shifts closer to this level. Consequently, the
SPE defect population increases causing a further enhancement of the
SPE rate. This suggests that the dopant concentration would be
`effectively' underestimated and that this underestimation increases
with dopant concentration or as band bending becomes less
pronounced. This may explain the concentration dependence of the SPE
defect level, $E^{\pm}$, in Fig.~\ref{figs/pndevsconc} that was found
through fitting our SPE data.

\section{Conclusion}
\label{conclusions}
Dopant-enhanced SPE has been measured for buried $a$-Si layers doped
with P, B or Al over the concentration range $\rm 1- 30\times 10^{19}
cm^{-3}$ and compared to As-enhanced SPE data published by McCallum in
Ref.~\cite{NIMB:mccallum}. The GFLS model was extended by seeking the
best values for temperature and concentration dependences of the
parameters involved. A theoretical calculation of the Fermi level for
extrinsic and degenerately doped Si fully justified the use of
degenerate semiconductor statistics in the concentration and
temperature range considered. Although there are relatively large
differences in $N_{i}$ values predicted using various
parameterisations we found that these only contribute a 3\% error in
$E^{\pm}$ values. Values of the energy level and the degeneracy of the
defect level responsible for SPE obtained from our fits were:
As($E_{c}-E^{-}=0.16$, $g=0.53$), P($E_{c}-E^{-}=0.23$, $g=0.25$),
B($E^{+}-E_{v}=0.17$, $g=1.5$) and Al($E^{+}-E_{v}=0.-0.08$,
$g=0.16$). Apart from the Al data which showed anomalous SPE regrowth
behaviour these values are remarkably similar to what one might expect
for a DB type defect despite the complexity of the fitting procedure.

The GFLS model was extended to consider band bending at the $c$-$a$
interface but the lack of a complete description of $a$-Si prevented
the extraction of any useful data at this time. However, it was
reasoned that band bending could explain the concentration dependence
of the defect level that was observed.

The calculation of the Fermi level in $c$-Si for SPE studies could be
made more self-consistent if the various parameterisations in the
literature are avoided in favour of Monte-Carlo simulations from first
principles. Advanced device simulations are often performed with the
aid of programs such as TCAD.\cite{tcad} This work would also benefit
from a similar calculation of the $c$-$a$ interface growth process in
order to link the band gap states reported in this paper with
particular defects. Designing an independent experiment to support the
results is difficult. However, the high quality of the data and fits
and the extensive parameter review presented here should serve as a
good starting point for such calculations.

\begin{acknowledgments}
The Department of Electronic Materials Engineering at the Australian
National University is acknowledged for their support by providing
access to SIMS and ion implanting facilities.
\end{acknowledgments}


\appendix*
\section{}

\subsection{Fermi Level Equations}
\label{sec:fermi}
The carrier concentrations in the conduction and valence bands in an
intrinsic semiconductor are commonly given by\cite{book:sze}

\begin{subequations}
\label{conc}
\begin{equation}
\label{electronconc}
n_{e}=N_{c} \exp\Bigl(\frac{E_{f}-E_{c}}{kT}\Bigr)
\end{equation}
and
\begin{equation}
\label{holeconc}
n_{h}=N_{v} \exp\Bigl(\frac{E_{v}-E_{f}}{kT}\Bigr)
\end{equation}
\end{subequations}

\noindent where $N_{c}$ and $N_{v}$ are the effective DOS in the
conduction and valence bands, respectively. For an intrinsic
semiconductor these concentrations are equal so these equations can be solved for the Fermi level in intrinsic Si, $E_{{fi}}$ given that energy levels of the conduction and valence band edges, $E_{c}$ and $E_{v}$, and the associated effective DOS are known.


Likewise, Eq.\ref{conc} can be solved for the Fermi level for an extrinsic semiconductor. For an $n$-type semiconductor, if the donor concentration, $N_{d}$ is large compared to the intrinsic carrier concentration, $n_{i}$ then it is a reasonable approximation to set the carrier concentration, $n_{e}$ equal to the ionized donor concentration.

However, for highly doped semiconductors Eq.~\ref{electronconc} is no longer valid as the expressions for the carrier concentrations are based on classical approximations to the Fermi distribution. These approximations deviate significantly from the Fermi distribution once the Fermi level lies
within 3$kT$ of the band edges. In this regime degenerate
semiconductor statistics must be used. Thus, Eq.~\ref{electronconc} becomes

\begin{equation} 
\label{nedegenerate}
n_{e}=\frac{2N_{c}}{\sqrt{\pi}}\mathcal{F}_{1/2}\Bigl(\frac{E_{f}-E_{c}}{kT}\Bigr)  
\end{equation} 

\noindent where $\mathcal F_{1/2}()$ is the Fermi-Dirac integral. The most accurate approximation to this intractable integral is the Bednarczyk approximation with an error of less than $0.3787\%$.\cite{PL:bednarczyk} We can use this approximation to numerically solve the charge neutrality condition,

\begin{equation} 
\label{chargedegenerate}
n_{e}-n_{h}=[N_{d}^{+}]-[N_{a}^{-}].
\end{equation}

For an $n$-type semiconductor we assume that $[N_{a}^{-}]=0$. The
concentration of charged donor ions, $[N_{d}^{+}]$, is given by the
Fermi-Dirac weighting function,

\begin{equation}
\label{conc:nd} 
[N_{d}^{+}]=\frac{N_{d}}{1+g \exp((E_{f}-E_{d})/kT)},
\end{equation}

\noindent where $E_{d}$ is the energy level that the charged donor ions introduce into the band gap. According to Sze and Irvin $E_{d}$ has values of $(E_{c}-49)$, and $(E_{c}-44)$~meV for As and P and acceptor energy levels of $(E_{v}+45)$ and $(E_{v}+57)$~meV for B and Al,
respectively.\cite{SSE:sze} The degeneracy factor, $g$ is equal to 2
for donor levels and 4 for acceptor levels.\cite{book:sze} In the
analysis that follows we assume that every implanted dopant atom is
electrically active and hence has the opportunity to become
ionized. However, it has been reported that a saturation of the SPE
regrowth rate is reached when the dopant concentrations of As, P or B
exceed their respective solid solubility limits.\cite{book:olson} In
this high concentration regime a fraction of the implanted ions do not
become electrically active. For the dopants analysed in our study this
limit generally represents the upper boundary of concentrations
examined.

The concentration of holes in an $n$-type semiconductor can be
determined with Boltzmann statistics as per Eq.~\ref{holeconc} since
the Fermi level is far from the valence band edge.  Eq.~\ref{chargedegenerate} for an $n$-type semiconductor then becomes

\begin{eqnarray}
\label{charge2}
&&\frac{2N_{c}}{\sqrt{\pi}}\mathcal{F}_{1/2}\Bigl(\frac{E_{f}-E_{c}}{kT}\Bigr)\nonumber\\
&&=\frac{N_{d}}{1+2 \exp((E_{f}-E_{d})/kT)}+N_{v} \exp\Bigl(\frac{E_{v}-E_{f}}{kT}\Bigr).
\end{eqnarray}

The Fermi level of a degenerately doped semiconductor can then be solved numerically and substituted into Eq.\ref{fit} if $E_{c}$, $E_{v}$ and their temperature and dopant concentration dependences are known. An outline of the parameters used in this calculation is presented in the next section.

\subsection{Si band structure}
\label{parameters}

There are three main sets of parameters that can be used to construct a picture of the band gap structure. These are: expressions outlined by Sze, commonly referred to as the $T^{3/2}$ model;\cite{book:sze} numerical relations compiled by Green;\cite{JAP:green} and Monte Carlo simulations.~\cite{PB:smith}. It is well known that the $T^{3/2}$ model is inaccurate even in the device operation temperature regime. Green's relations are valid only up to 500~K. However, these relations are often extrapolated for use in device simulations in the processing temperature regime given that measurements performed at elevated temperatures are lacking in the literature. Monte Carlo simulations do offer consistent and physical models but require sophisticated simulation software.

This section reviews our current understanding of the parameters used to describe the band structure of Si. It also aims to select a consistent and reliable parameter set for intrinsic and doped $c$-Si between 460 and 660$\rm ^{o}$C so that the expressions for the Fermi levels presented above can be calculated. Monte Carlo simulations are not considered at this time.

\subsubsection{Band Gap Narrowing}
The band gap width has a temperature dependence arising from the
dilation of the lattice. At elevated temperatures, electron-phonon
interactions also become important. Theory predicts that thermal band
gap narrowing (BGN) should be linear at high temperatures and
non-linear at low temperatures. This behaviour is well described by
the semi-empirical formula given by Varshni,\cite{physica:varshni}

\begin{equation}
\label{varshni}
E_{g}=E_{o}-\frac{\alpha T^{2}}{T+\beta}
\end{equation}

\noindent where $E_{o}$ is the energy gap at $T=0$~K. It is usually
assumed that this variation is distributed evenly between the
conduction and valence bands so that each is shifted by an amount
$E_{g}/2$ towards mid-gap. $\alpha$ and $\beta$ are fitting parameters
which have taken on a number of different values in the past depending
on the availability of data. Recently, Alex {\em et al.} have
performed photoluminescence experiments and determined the BGN in the temperature range 2 - 750~K.\cite{JAP:alex} They have
found $\alpha =4.9 \times 10^{-4}$~$\rm eV.K^{-1}$ and $\beta =655$~K
with $E_{g}=1.1692$~eV. Smith {\em et al.} have used these parameters to
model transition metal defect behaviour between 1100 and
1400~K.\cite{PB:smith} 

The band gap is also known to be reduced upon heavy doping (for a review see
Ref.~\cite{AEEP:jain}). The concentration dependence of BGN arises through the interaction of carriers created thermally with carriers introduced by the dopant and the dopant ion itself.\cite{JAP:lindefelt} An exact description of this narrowing has been controversial because optical and electrical measurements give different results with the latter yielding considerably higher BGN values. Klaassen has presented a unified apparent BGN function.\cite{SSE:klaassen} This formulation has brought together the disparate data from optical and electrical measurements by correcting the transport equations used in models to interpret electrical data with new accurate values of the intrinsic carrier concentration reported by Green.\cite{JAP:green} This formulation also agrees fairly well for dopant concentrations between $\rm 1 \times 10^{18}$ and $\rm 1 \times 10^{20} \; cm^{-3}$ to a recent theoretical model proposed by Schenk based on quantum mechanical principles using a full random phase approximation.\cite{JAP:schenk}

\subsubsection{Effective DOS and the Intrinsic Carrier Concentration}
\label{dosmass}

The effective DOS is used to calculate the concentration of carriers
in the conduction and valence bands in Eq.~\ref{conc}. Often, the
effective DOS is calculated assuming that the bands are parabolic
resulting in the $T^{3/2}$ model given by Sze\cite{book:sze}

\begin{eqnarray} 
\label{szedos}
N_{c} &= 2 (2 \pi m^{*}_{e} k T/h^{2})^{3/2} M_{c}\nonumber\\ 
& \\
N_{v} &= 2 (2 \pi m^{*}_{h} k T/h^{2})^{3/2} \nonumber \\ & \nonumber
\end{eqnarray} 

\noindent where $M_{c}$ is the number of equivalent minima in the
conduction band and $m^{*}_{e}$ and $m^{*}_{h}$ are the effective
electron and hole masses, respectively. To our knowledge there is no
experimental data for the effective mass or the effective DOS in the
processing temperature regime of interest in our SPE
measurements. Therefore, an extrapolation of fits to data collected at
lower temperatures is unavoidable.

The electron effective mass used by Green to calculate the effective DOS is valid up to a temperature of 300~K and was found to depend on the band gap width.\cite{JAP:green} Green's electron effective mass values compare well with results by Hensel {\em et al.} at a
temperature of 4.2~K.\cite{PR:hensel} Green's values at room temperature are lower than the commonly used values of Barber.\cite{SSE:barber} However, the expected weak temperature dependence ensured that values would be accurate to within a few percent up to a temperature of 500~K.

The hole effective mass reported by Green is much greater than that
reported by Barber and for temperatures above $\sim$230~K Green's
relations suggest that the hole effective mass becomes greater than
the electron effective mass. However, this is justified by more recent
and rigorous calculations performed by Humphreys,\cite{JPC:humphreys} Madarasz {\em et al.},\cite{JAP:madarasz} and indium ionisation
data by Parker\cite{JAP:parker} for temperatures up to 500~K.

The intrinsic carrier concentration has been determined experimentally by Morin and Maita for temperatures between 10 and 1100~K.\cite{PR:morin} We can therefore look for a consistent set of
parameters that follow the trend in the Morin and Maita data by considering the law of mass action which relates the effective DOS and the intrinsic carrier concentration, $n_{i}$:

\begin{equation}
\label{massaction2}
n_{e}n_{h}=n_{i}^2=N_{c}N_{v} \; e^{-E_{g}/kT}.
\end{equation} 

\begin{figure}[!t]
\begin{center}
\rotatebox{180}{\includegraphics[height= 15cm]{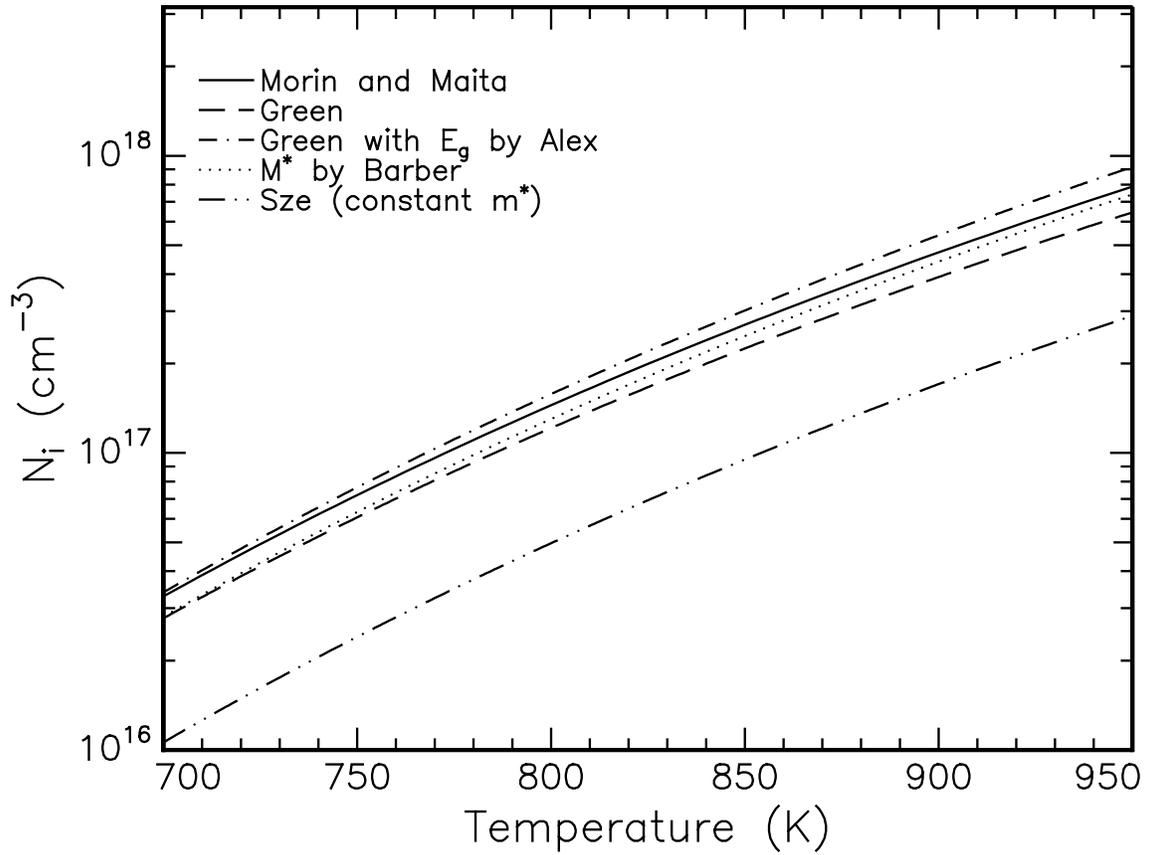}}
\end{center}
\caption[Intrinsic carrier concentration as a function of
temperature]{The intrinsic carrier concentration as a function of
temperature calculated from relations by Green\cite{JAP:green} and
Barber\cite{SSE:barber}. Green's relations are also expressed using the
thermal BGN equation determined by Alex.\cite{JAP:alex}}
\label{figs/Ni.ps}
\end{figure}

Figure~\ref{figs/Ni.ps} shows the variation of the intrinsic carrier
concentration according to several authors as a function of
temperature over the range of our SPE data. The intrinsic carrier
concentration by Morin and Maita is shown as a solid line. If we
assume that the effective mass has no temperature dependence and that
the effective DOS has only a $T^{3/2}$ dependence then we obtain the
solitary curve (dot-dot-dash curve) in Fig.~\ref{figs/Ni.ps}.

We have chosen to use the Varshni equation (Eq.~\ref{varshni}) with parameters reported by Alex\cite{JAP:alex} and the effective electron mass reported by Green {\em et al.}.\cite{JAP:green,JAP:lang} This results in a 3\% variation relative to the Morin and Maita data.  

Furthermore, we have also chosen to ignore the concentration dependence of the effective mass. The DOS of the conduction band edge is large in Si and the non-parabolicity is small so a change in effective mass is expected to only occur for extremely high fluences. Generally, the free carrier effective mass may increase significantly for concentrations in excess of $\rm 1 \times 10^{21} \; cm^{-3}$.\cite{ssc:miyao,PSS:slaoui} However, no empirical formulations of the concentration dependence exists in the literature and the highest concentration used in the experiments reported in this paper is $\rm 3\times 10^{20} \; cm^{3}$.

\subsection{Fermi Level Calculations}
\begin{figure}[!ht]
\begin{center}
\rotatebox{180}{\includegraphics[height= 15cm]{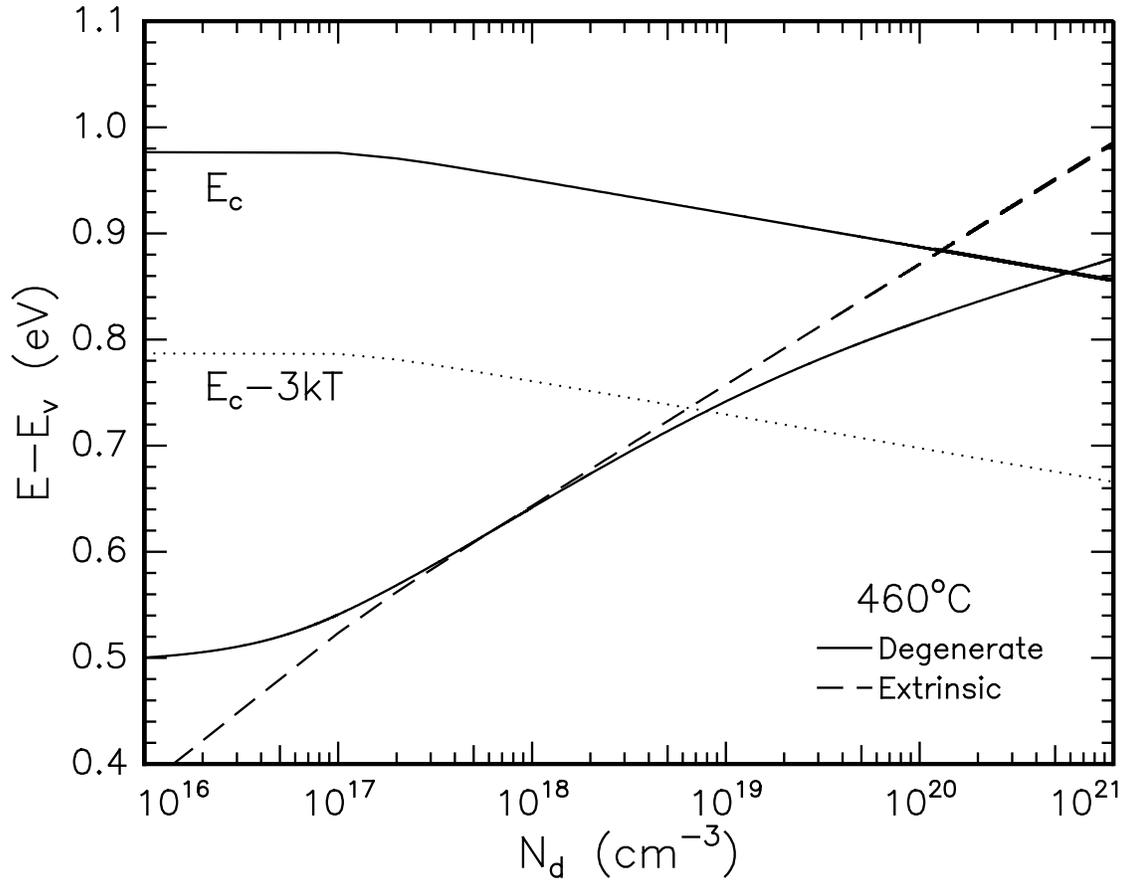}}%
\end{center}
\caption[Fermi level as a function of donor
concentration]{\label{figs/degenwindowbgn} The Fermi level as a
  function of donor concentration calculated by solving the electrical
  neutrality condition for a degenerate semiconductor
  (Eq.~\ref{charge2}). The dashed line represents the Fermi level
  calculated using non-degenerate semiconductor statistics
  (from Eq.~\ref{conc}). The dotted line represents a 3$kT$ window
  beyond which a degenerate approach must be taken.}
\end{figure}

Figure~\ref{figs/degenwindowbgn} shows the concentration at which degenerate semiconductor statistics become important. Here the Fermi level was calculated as a function of the dopant concentration using Eq.~\ref{charge2} at a temperature of 460$^{o}$C. This is the lowest temperature used in these experiments and, for a constant
concentration, the Fermi level will be closest to the band edge at this temperature. Above a dopant concentration of $\rm 7.3 \times 10^{18}\; cm^{-3}$ the Fermi level crosses over the 3$kT$ window and into the degenerate regime. This is true for both As and P donor
impurities as their defect levels lie relatively close together and are expected to have a similar effect on the Fermi level position.

Fig.~\ref{figs/degenwindowbgn} also shows the Fermi level of an
extrinsic semiconductor calculated using
Eq.~\ref{conc}. Both Fermi levels agree within a dopant
concentration range of about $\rm 1 \times 10^{17}-1 \times 10^{19}\;
cm^{-3}$. In the lower concentration range ($<\rm 1 \times 10^{17} cm^{-3}$)
the approximation that $n_{e}\simeq N_{d}$ is no longer appropriate as
carriers generated thermally will dominate the electrical properties
of the semiconductor. In the high concentration regime ($>\rm 1 \times
10^{19}\; cm^{-3}$) the classical distribution function cannot be
used.

For a $p$-type semiconductor the concentration at which the Fermi
level crosses the 3$kT$ window will be different. This is mainly a result of
the valence band effective DOS being about 40\% greater than the conduction band effective DOS as calculated with Green's relations. Therefore, we might expect the dopants to have less effect on the Fermi level position. Using the same method as above the concentration at which a boron doped semiconductor becomes degenerate is $\rm 1.03 \times 10^{19} cm^{-3}$. For Al, having a deeper ionization level in the band gap, the concentration is
$\rm 1.09 \times 10^{19} cm^{-3}$.

Dopant concentrations at which dopant-enhanced SPE is observable are
generally above $\rm 1 \times 10^{19} \; cm^{-3}$ so using a
degenerate approach to calculate the Fermi level is totally justified
for all dopant-enhanced SPE studies.


\newpage

\newpage

\newpage

\newpage

\newpage

\newpage

\newpage

\newpage

\newpage

\newpage

\newpage

\newpage

\begin{figure}
\rotatebox{0}{\includegraphics[height= 17cm]{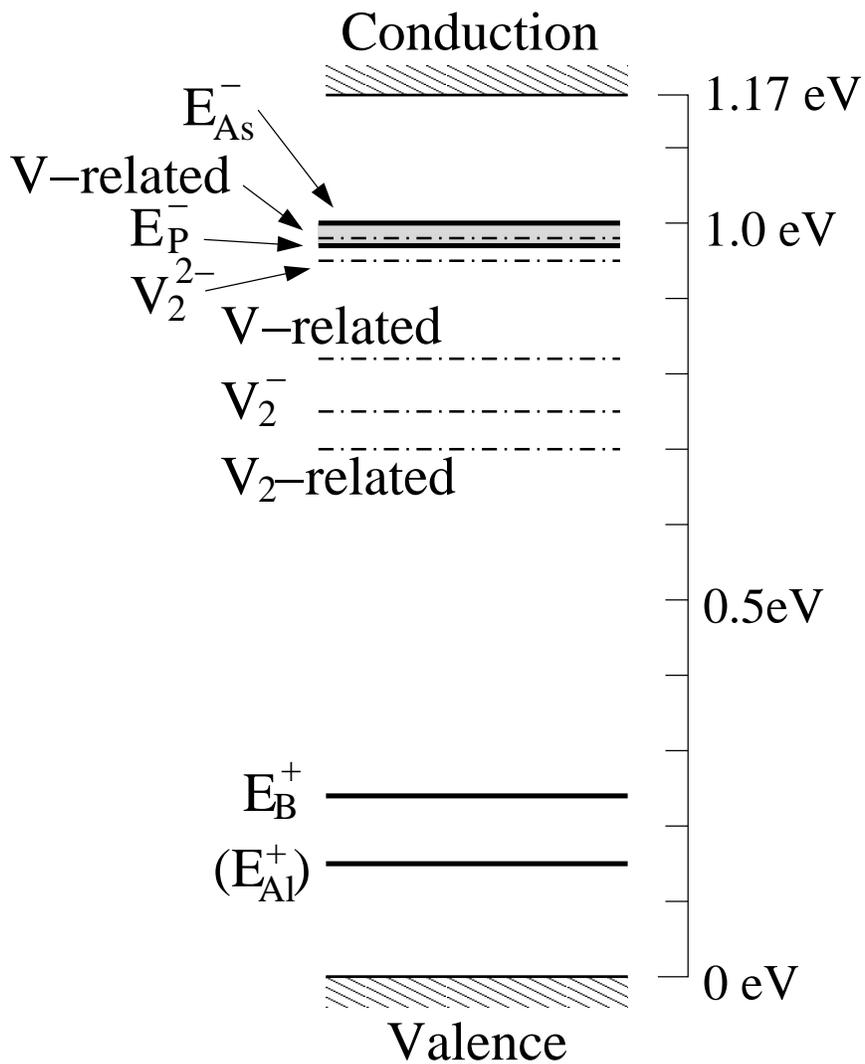}}
\caption[Band gap energy level diagram of Si with vacancy and SPE
defect levels]{Band gap of Si at room temperature with the energy
levels associated with the SPE defect identified with the GFLS
model. The energy levels of vacancy related defects are shown for
comparison.~\cite{JAP:palmetshofer,JAP:huppi} Levels are referenced to 
the edge of the valence band.}
\label{band}
\end{figure}

\newpage

\newpage

\newpage

\end{document}